\pdfoutput=1
\documentclass[journal]{IEEEtran}
\usepackage{microtype}

\usepackage[T1]{fontenc}
\usepackage{amsmath,amssymb,amsthm,booktabs,array,graphicx,url,xcolor,cite}
\usepackage{algorithm,algorithmic}
\usepackage{hyperref}
\AtBeginDocument{%
  \setlength{\abovedisplayskip}{2.2pt plus 0.4pt minus 0.6pt}%
  \setlength{\belowdisplayskip}{2.2pt plus 0.4pt minus 0.6pt}%
  \setlength{\abovedisplayshortskip}{1.5pt plus 0.5pt}%
  \setlength{\belowdisplayshortskip}{1.5pt plus 0.5pt}%
}
\hypersetup{colorlinks=true,linkcolor=black,citecolor=black,urlcolor=blue,pdftitle={Energy-Efficient Mobile Embodied Agent Network (MEAN) over Wireless Communications},pdfauthor={Zhicheng Liu, Chen Zhu, Zhouxiang Zhao, Jiaxiang Wang, Zhaohui Yang, Mohammad Shikh-Bahaei, and Zhaoyang Zhang}}
\newtheorem{proposition}{Proposition}
\newtheorem{theorem}{Theorem}
\newtheorem{lemma}{Lemma}
\newtheorem{remark}{Remark}
\begin{document}
\title{Energy-Efficient Mobile Embodied Agent Network (MEAN) over Wireless Communications}
\author{Zhicheng~Liu, Chen~Zhu, Zhouxiang~Zhao, Jiaxiang~Wang, Zhaohui~Yang, Mohammad~Shikh-Bahaei,~\IEEEmembership{Senior~Member,~IEEE,} and~Zhaoyang~Zhang,~\IEEEmembership{Fellow,~IEEE}%
\thanks{\emergencystretch=1em Z. Liu, Z. Zhao, Z. Yang, and Z. Zhang are with the College of Information Science and Electronic Engineering, Zhejiang University, Hangzhou 310027, China (e-mail: \{liu\_zc,\allowbreak zhouxiangzhao,\allowbreak yang\_zhaohui,\allowbreak ning\_ming\}@zju.edu.cn).}%
\thanks{C. Zhu is with the School of Communication Engineering, Hangzhou Dianzi University, Hangzhou 310018, China (e-mail: zhuc@zju.edu.cn).}%
\thanks{\emergencystretch=1em J. Wang and M. Shikh-Bahaei are with the Department of Engineering, King's College London, London WC2R 2LS, U.K. (e-mail: \{jiaxiang.wang,\allowbreak m.sbahaei\}@kcl.ac.uk).}\vspace{-10pt}}
\markboth{}{}%
\maketitle

\begin{abstract}
In this paper, we investigate energy-efficient joint trajectory, semantic compression, and transmit power optimization for a mobile embodied agent network (MEAN) over wireless communications. In the considered system, each robot senses an assigned target and relocates toward higher channel gain to transmit its compressed update to a multi-antenna base station (BS). During relocation, shared knowledge-graph compression overlaps with travel and shortens transmission duration at the expense of computation energy and, under lossy pruning, reconstruction fidelity. To ensure target-level fairness, we maximize the minimum time-averaged quality--freshness utility under sensing, service quality, and per-robot battery constraints. For a symmetric binary Markov source under exogenous latest-valid-packet estimation with calibrated symmetric conditionally independent errors, this objective is equivalent to minimizing worst-target Bayesian risk. For a fixed plan, we propose an alternating algorithm with a seven-regime compression optimality law at stationary anchors, closed-form quality-preserving pruning, and monotonic objective convergence of retained iterates. Numerical results show that under equal battery budgets, the proposed algorithm improves energy efficiency with $28.2\%$ higher utility and $17.2\%$ lower swarm energy than outer-boundary waypoints at nominal load, and outperforms co-located operation by $9.2\%$ under heavy payloads and fade-margin planning by $24.6\%$ under shadowing.
\end{abstract}
\IEEEpeerreviewmaketitle

\section{Introduction}
\label{sec:intro}

\IEEEPARstart{T}{he} mobile embodied agent network (MEAN) connects autonomous robots to wireless access points and edge processors in sixth-generation (6G) networks~\cite{li2025embodied,letaief2019roadmap,saad2020vision,gunduz2023beyond,zhao2025energy}. During continuous service, each robot revisits assigned ground targets so the base station (BS) maintains accurate target-state estimates. Semantic communication transmits task-relevant features in place of raw sensory streams~\cite{xie2021deep,gunduz2023beyond}. Specifically, a robot extracts entity--relation triples from visual or light detection and ranging (LiDAR) data while the BS infers predictable triples via a shared probabilistic knowledge graph~\cite{zhao2025energy,zhu2025scene}. Stronger compression shortens wireless transmission duration at higher computation delay and energy. Waypoint control governs sensing accuracy, wireless channel gain, and propulsion energy~\cite{mei2006deployment,zeng2017energy,zeng2019accessing,zeng2021toward,zhu2024movable}, while evolving target states erode observation value before BS reconstruction~\cite{yates2021age,sun2017update,maatouk2020age,kountouris2021semantics}.

Prior mobile wireless studies examined how trajectories and schedules shape energy, delay, and throughput~\cite{li2025embodied,mei2006deployment,zeng2017energy,zeng2019accessing,wu2018joint}. Robot deployment studies characterize the trade-off between tour energy and delay~\cite{mei2006deployment}. Two-dimensional (2D) uncrewed aerial vehicle (UAV) trajectory optimization reduces propulsion energy or age of information (AoI)~\cite{zeng2017energy,zeng2019accessing,wu2018joint,huang2025learning}, and multi-source scheduling governs AoI over shared wireless links~\cite{kadota2018scheduling}. Jiang et al.\ separated sensing and transmission positions to reduce AoI~\cite{jiang2024age}. Liu et al.\ jointly optimized UAV trajectories, onboard computing, and wireless transmission under energy budgets~\cite{liu2025joint}. The above works~\cite{mei2006deployment,zeng2017energy,zeng2019accessing,wu2018joint,huang2025learning,kadota2018scheduling,jiang2024age,liu2025joint} treat updates as fixed bit payloads without optimizing compression during travel. Our conference work~\cite{liu2026mobility} optimized travel, compression, and power for one mobile agent, whereas under concurrent multi-robot transmissions each update's utility depends on arrivals from other visits.

Movable-antenna and fluid-antenna systems also use spatial displacement to improve wireless channels~\cite{zhu2024movable,ma2024mimo,wong2021fluid}. Mobile edge computing balances wireless transmission and central processing unit (CPU) energy~\cite{mao2017survey,you2017energy}, while iterative clustering allocates multi-robot tasks~\cite{martin2026collaborative}. For a fixed active transmitter set, block successive minimization~\cite{razaviyayn2013unified}, majorization and minimization~\cite{sun2017majorization}, fractional programming~\cite{shen2018fractional}, weighted minimum mean-square error (MMSE)~\cite{shi2011iteratively}, and utility decomposition~\cite{palomar2006tutorial} provide tractable interference management. In multi-robot monitoring, shifting a robot's schedule alters active transmitters and reshapes multi-user linear minimum mean-square error (LMMSE) signal-to-interference-plus-noise ratios (SINRs)~\cite{bjornson2015optimal,tse1999linear}.

Semantic communication lowers bandwidth demand via task-relevant extraction~\cite{gunduz2023beyond,luo2022semantic,qin2023generalized,yang2023semantic}, including deep joint source and channel coding~\cite{bourtsoulatze2019deep}, nonlinear transform coding~\cite{dai2022nonlinear}, DeepSC~\cite{xie2021deep}, multi-user encoding~\cite{xie2022task}, and offloading~\cite{ji2024resource}. Recent UAV semantic works optimize trajectories and resources~\cite{wu2026joint,sun2026joint,huang2025joint}, with~\cite{wu2026joint,huang2025joint} modeling AoI or freshness. Knowledge graphs~\cite{zhou2024cognitive} and probabilistic semantic communication (PSCom)~\cite{zhao2025energy,zhu2025scene} enable continuous compression with convex piecewise-affine CPU workloads~\cite{zhao2025energy}, yet decoupled sensing and transmission waypoints with in-transit compression remain unaddressed. Optimal sensing favors target proximity, whereas optimal transmission favors unobstructed BS links and low interference. When transit exceeds compression delay, saved transmission duration must exceed residual travel to advance delivery under shared battery budgets.

Motivated by these observations, we investigate joint waypoint, semantic compression, transmit power, and scheduling optimization for a freshness-aware MEAN. Our main contributions are summarized below.
\begin{itemize}\setlength{\itemsep}{-0.6pt}
\item First, we establish a multi-robot MEAN model with decoupled sensing and transmission waypoints, knowledge-graph compression during travel, location-dependent channel knowledge map (CKM) and spatially correlated multipath channels, and asynchronous multi-user LMMSE reception under imperfect channel state information (CSI), including co-located operation as a special case.
\item To ensure target-level fairness, we formulate a max--min fairness problem maximizing the minimum time-averaged quality--freshness utility across targets under tightened sensing areas, multi-scenario service quality, and per-robot battery constraints. Under exogenous latest-valid-packet estimation with calibrated symmetric conditionally independent errors for a symmetric binary Markov source, we prove equivalence to minimizing worst-target Bayesian estimation risk.
\item For a fixed monitoring plan, we propose a block-coordinate descent (BCD) and successive convex approximation (SCA) algorithm for the continuous subproblem of the non-convex mixed-integer nonlinear program (MINLP). At a stationary anchor with finite duals, Karush--Kuhn--Tucker (KKT) conditions of the compression subproblem characterize seven optimal regimes, with a closed form under quality-preserving pruning, while second-order cone programming (SOCP) updates the waypoints. Retained iterates achieve monotonic objective convergence alongside analytical battery, spatial degrees-of-freedom (DoF), and utility bounds.
\item Simulation results show that under equal battery budgets, the proposed algorithm improves energy efficiency with $28.2\%$ higher utility and $17.2\%$ lower swarm energy than outer-boundary waypoints at nominal load, and outperforms co-located operation by $9.2\%$ in the payload sweep and fade-margin planning by $24.6\%$ under obstacle shadowing.
\end{itemize}

The rest of our paper proceeds as follows. Section~\ref{sec:model} introduces the system model and formulates the problem, Section~\ref{sec:algorithm} develops and analyzes the proposed algorithm, Section~\ref{sec:results} provides simulation results, and Section~\ref{sec:conclusion} concludes this paper. Throughout this paper, $(\cdot)^\top$, $(\cdot)^H$, $\|\cdot\|$, and $\|\cdot\|_\infty$ denote transpose, conjugate transpose, Euclidean norm, and $\ell_\infty$ norm, respectively, $[u]_+=\max\{u,0\}$, and boldface letters denote stacked variable collections.

\section{System Model and Problem Formulation}
\label{sec:model}

\begin{figure*}[t]
\centering
\includegraphics[trim=12bp 6bp 13bp 3bp,clip,width=.91\textwidth]{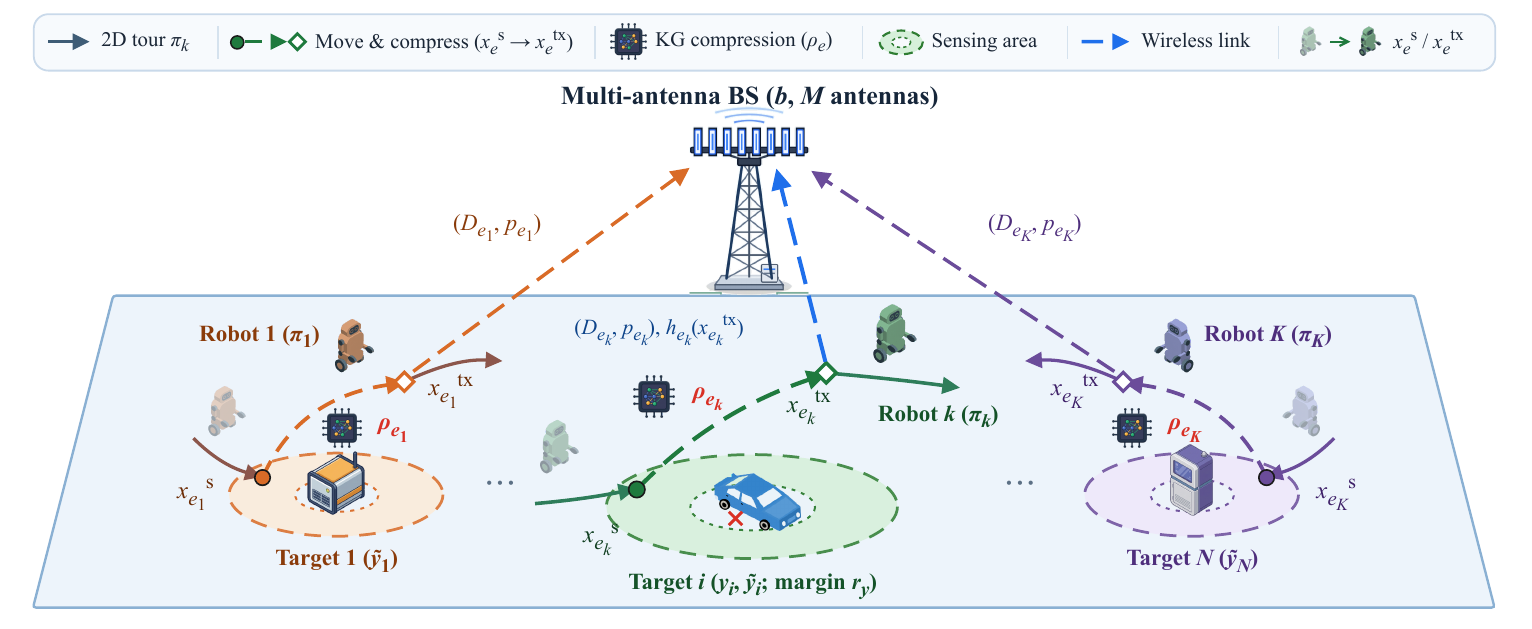}
\caption{Mobile robots sense ground targets, compress observations during travel, and transmit to a multi-antenna BS.}
\label{fig:system}
\end{figure*}

As shown in Fig.~\ref{fig:system}, we consider a MEAN where $K$ single-antenna mobile robots monitor $N$ stationary ground targets and transmit compressed semantic updates to an $M$-antenna uniform linear array BS. Let $\mathcal{K}=\{1,\ldots,K\}$ and $\mathcal{I}=\{1,\ldots,N\}$ denote the robot and target sets, respectively. Each robot follows an assigned monitoring tour~\cite{mei2006deployment}, compresses observations via a shared probabilistic knowledge graph~\cite{zhao2025energy,zhu2025scene}, and transmits concurrently under LMMSE reception. Table~\ref{tab:notations} lists the main notations alphabetically.

\begin{table}[t]
\centering
\caption{Summary of Main Notations}
\label{tab:notations}
\footnotesize
\setlength{\tabcolsep}{3.5pt}
\renewcommand{\arraystretch}{0.88}
\begin{tabular}{@{}p{0.34\columnwidth}>{\raggedright\arraybackslash}p{0.61\columnwidth}@{}}
\toprule
\textbf{Notation} & \textbf{Physical Meaning}\\
\midrule
$b,b_i,B_k$ & BS coordinate, packet header, robot-$k$ battery budget\\
$c_e,d_e,d_{\min,i},d_{\max,i}$ & Workload epigraph, BS completion time, sensing bounds\\
$D_i,D_e$ & Target-$i$ raw feature size, task-$e$ compressed payload\\
$E_k^{\mathrm{cpu}},E_k^{\mathrm{mov}},E_k^{\mathrm{rad}}$ & Onboard CPU, propulsion, and wireless energies\\
$\mathcal{E},F_k,\mathcal{I},\mathcal{K}$ & Task set, robot-$k$ CPU clock, target and robot sets\\
$f_e,g_e,h_e,\hat{h}_e^\omega$ & Finish/capture times; nominal and est.\ channels\\
$L_k(\mathbf{x}),p_e,q_e^{\mathrm{sp}}$ & Robot-$k$ tour length, transmit power, spatial quality\\
$r_y,R_e,\bar{\mathcal{R}}_i$ & Radial safety margin, achievable rate, Bayesian risk\\
$s_e,S_e^\omega,T_e^{\mathrm{cpu}}$ & Transmission start time, bit service, onboard CPU delay\\
$U_i,U,w_e$ & Target-$i$ utility, minimum utility, LMMSE combiner\\
$x_e^{\mathrm{s}},x_e^{\mathrm{tx}},y_i,\tilde{y}_i$ & Sensing/transmission waypoints; true/estimated target\\
$\gamma_e,\Gamma_i(\rho_e),\epsilon_k,\zeta$ & SINR, CPU workload in cycles, J/cycle, service reserve\\
$\pi_k,\Pi,\rho_e,\rho_{\min,i}$ & Robot-$k$ tour, monitoring plan, compression ratio/floor\\
$\tau_i,\psi_i(\rho_e)$ & Freshness time constant, semantic fidelity function\\
$\widetilde{\Phi}_e^{\mathrm{net}},\Psi_e^{\mathrm{cpu}},\tilde{\vartheta}_e^0$ & Net compression, CPU, and dual price ratios\\
\bottomrule
\end{tabular}
\vspace{-2pt}
\end{table}

\subsection{Mobility and Semantic Models}
The mission runs over $[0,H]$ of duration $H>0$~s. Each target $i\in\mathcal{I}$ resides at true coordinate $y_i\in\mathcal{X}$ within a bounded convex workspace $\mathcal{X}\subset\mathbb{R}^2$ admitting an SOCP representation. Over $[0,H]$, the swarm executes $P$ tasks indexed by $\mathcal{E}=\{1,\ldots,P\}$. Task $e\in\mathcal{E}$ is assigned to robot $k_e\in\mathcal{K}$ for target $i_e\in\mathcal{I}$ with decoupled waypoints $x_e=(x_e^{\mathrm{s}},x_e^{\mathrm{tx}})\in\mathcal{X}^2$, where $x_e^{\mathrm{s}}$ and $x_e^{\mathrm{tx}}$ denote sensing and transmission waypoints, respectively, and $[s_e,f_e)\subseteq[0,H]$ with $s_e\le f_e$ is the transmission interval from start $s_e$ to finish $f_e$. Robot $k_e$ observes target $i_e$ at $x_e^{\mathrm{s}}$, moves to $x_e^{\mathrm{tx}}$ while compressing onboard, and stays stationary over $[s_e,f_e)$, where $x_e^{\mathrm{tx}}=x_e^{\mathrm{s}}$ recovers co-located operation. With robot $k$'s ordered sequence $\pi_k=(e_{k,1},\ldots,e_{k,|\pi_k|})$, the monitoring plan $\Pi=(P,\{i_e\}_{e=1}^P,\{\pi_k\}_{k\in\mathcal{K}})$ partitions $\mathcal{E}$ and determines $k_e$ for each $e\in\mathcal{E}$.

Let $d_{\min,i}>0$ and $d_{\max,i}>d_{\min,i}$ denote target $i$'s minimum stand-off distance and maximum sensing range, respectively. To avoid collision and occlusion while maintaining coverage, $x_e^{\mathrm{s}}$ lies in the ring around $y_{i_e}$ and $x_e^{\mathrm{tx}}\in\mathcal{X}$~\cite{mei2006deployment,zeng2019accessing,wu2018joint}.
\begin{equation}
x_e^{\mathrm{s}},x_e^{\mathrm{tx}}\in\mathcal{X},\qquad d_{\min,i_e}\le\|x_e^{\mathrm{s}}-y_{i_e}\|\le d_{\max,i_e}.
\label{eq:sensing_area}
\end{equation}
In practice, the planner only observes noisy target position estimates $\tilde{y}_i\in\mathbb{R}^2$ subject to independent zero-mean Gaussian per-axis errors with standard deviation $\sigma_y>0$. Hence $\|\tilde{y}_i-y_i\|$ follows a Rayleigh distribution with tail probability $\mathbb{P}(\|\tilde{y}_i-y_i\|>r_y)=\exp(-r_y^2/(2\sigma_y^2))$ for any margin $r_y>0$. Enforcing simultaneous sensing validity $\mathbb{P}(\max_{i\in\mathcal{I}}\|\tilde{y}_i-y_i\|>r_y)\le\alpha_y$ for a target outage tolerance $\alpha_y\in(0,1)$ via the union bound yields the radial safety margin
\begin{equation}
r_y=\sigma_y\sqrt{2\ln(N/\alpha_y)},
\label{eq:buffer}
\end{equation}
under $2r_y<d_{\max,i}-d_{\min,i}$ with non-empty intersection in $\mathcal{X}$. Substituting~\eqref{eq:buffer} into~\eqref{eq:sensing_area} yields the tightened sensing constraint
\begin{equation}
d_{\min,i_e}+r_y\le\|x_e^{\mathrm{s}}-\tilde{y}_{i_e}\|\le d_{\max,i_e}-r_y.
\label{eq:tight_sensing}
\end{equation}
By the triangle inequality,~\eqref{eq:tight_sensing} guarantees~\eqref{eq:sensing_area} simultaneously across all $i\in\mathcal{I}$ whenever $\max_{i\in\mathcal{I}}\|\tilde{y}_i-y_i\|\le r_y$, which holds with probability at least $1-\alpha_y$.

At $x_e^{\mathrm{s}}$, robot $k_e$ extracts $D_{i_e}>0$ raw feature bits as entity--relation triples~\cite{zhou2024cognitive,zhao2025energy,zhu2025scene}, and the BS infers predictable triples via the shared knowledge graph~\cite{zhao2025energy,zhu2025scene}. Let $\rho_e\in[\rho_{\min,i_e},1]$ denote the compression ratio, $\rho_{\min,i}\in(0,1)$ its floor, and $b_i\ge 0$ the header length in bits. The compressed bit payload $D_e$ is
\begin{equation}
\rho_{\min,i_e}\le\rho_e\le 1,\qquad D_e=b_{i_e}+\rho_e D_{i_e}.
\label{eq:payload}
\end{equation}
At $\rho_e=1$, all extracted triples are transmitted; reducing $\rho_e$ omits predictable triples and yields reconstruction fidelity $\psi_{i_e}(\rho_e)\in(0,1]$ at the BS.

Evaluating conditional probability tables over the shared knowledge graph~\cite{zhao2025energy,zhu2025scene} induces an $A$-stage continuous convex piecewise-affine CPU workload $\Gamma_i(\rho_e)$ in cycles with $A\ge 1$. Let $A_{ia}<0$ and $B_{ia}\in\mathbb{R}$ denote the slope and intercept of stage $a\in\{1,\ldots,A\}$. Then
\begin{equation}
\Gamma_i(\rho_e)=\max_{a\in\{1,\ldots,A\}}\{A_{ia}\rho_e+B_{ia}\}\ge 0.
\label{eq:workload}
\end{equation}
Introducing an auxiliary epigraph variable $c_e\ge 0$ for task $e$ linearizes~\eqref{eq:workload} as
\begin{equation}
c_e\ge 0,\qquad c_e\ge A_{i_e a}\rho_e+B_{i_e a},\quad a=1,\ldots,A.
\label{eq:epigraph}
\end{equation}
In the three-stage PSCom model with $A=3$~\cite{zhao2025energy,zhu2025scene}, breakpoints $\rho_{\min,i}<\varrho_2<\varrho_1<1$ separate marginal filtering, conditional-table pruning, and subgraph matching as $\rho_e$ decreases, with slopes $A_{ia}=-\varsigma_a c_i$ given by scale $c_i>0$ and $0<\varsigma_1<\varsigma_2<\varsigma_3$. Eliminating additional bits thus costs more CPU cycles at smaller $\rho_e$.

Let $F_k>0$ denote robot $k$'s CPU clock frequency, $t_{0,i}>0$ and $E_{0,i}>0$ the baseline extraction delay and energy, and $\epsilon_k=\kappa_{\mathrm{cpu}}F_k^2$ with $\kappa_{\mathrm{cpu}}>0$ the complementary metal--oxide--semiconductor (CMOS) dynamic energy per cycle~\cite{mao2017survey,you2017energy}. Task $e$'s CPU delay $T_e^{\mathrm{cpu}}$ and energy $E_{\mathrm{pkt},e}^{\mathrm{cpu}}$ are
\begin{equation}
T_e^{\mathrm{cpu}}=t_{0,i_e}+\Gamma_{i_e}(\rho_e)/F_{k_e},\quad E_{\mathrm{pkt},e}^{\mathrm{cpu}}=E_{0,i_e}+\epsilon_{k_e}\Gamma_{i_e}(\rho_e).
\label{eq:cpu_energy}
\end{equation}

\subsection{Multi-Antenna Wireless Transmission Model}
The BS resides at $b\in\mathbb{R}^2$ with elevation $h_{\mathrm{BS}}>0$ and unit array-axis vector $a_{\mathrm{arr}}\in\mathbb{R}^2$ satisfying $\|a_{\mathrm{arr}}\|=1$. At $x_e^{\mathrm{tx}}$, robot $k_e$ transmits with power $p_e\in[0,p_{\max,k_e}]$ for $p_{\max,k}>0$. Its three-dimensional (3D) distance $r_e$ and directional cosine $u_e$ are given by
\begin{equation}
r_e=\sqrt{\|x_e^{\mathrm{tx}}-b\|^2+h_{\mathrm{BS}}^2},\qquad u_e=a_{\mathrm{arr}}^\top(x_e^{\mathrm{tx}}-b)/r_e.
\label{eq:geom_ula}
\end{equation}
Let $W>0$ be bandwidth, $N_0>0$ noise power spectral density, $F_n\ge 1$ noise figure, $\delta=d_{\mathrm{ant}}/\lambda>0$ the inter-element spacing $d_{\mathrm{ant}}$ normalized by carrier wavelength $\lambda$, $\mathrm{j}=\sqrt{-1}$, and $\bar{\beta}:\mathcal{X}\to(0,\infty)$ a continuously differentiable large-scale path-gain map. In homogeneous settings, $\bar{\beta}(x_e^{\mathrm{tx}})=\beta_0(r_e/r_0)^{-\nu}$ with reference gain $\beta_0>0$ at $r_0=1\,\text{m}$ and exponent $\nu\ge 2$; under a site-specific CKM~\cite{zeng2021toward}, $\bar{\beta}(x_e^{\mathrm{tx}})$ models smooth obstacle blockage and shadowing in Section~\ref{subsec:ckm_eval}. Define $a_{\mathrm{st}}(u_e)\in\mathbb{C}^{M\times 1}$ by $[a_{\mathrm{st}}(u_e)]_{m_{\mathrm{a}}}=e^{-\mathrm{j}2\pi\delta m_{\mathrm{a}}u_e}$ for $m_{\mathrm{a}}\in\{0,\ldots,M-1\}$. From~\eqref{eq:geom_ula}, $h_e(x_e^{\mathrm{tx}})\in\mathbb{C}^{M\times 1}$, written $h_e$, is~\cite{zeng2019accessing}
\begin{equation}
h_e(x_e^{\mathrm{tx}})=\sqrt{\frac{\bar{\beta}(x_e^{\mathrm{tx}})}{M N_0 W F_n}}\,a_{\mathrm{st}}(u_e).
\label{eq:channel}
\end{equation}
Let $I_q$ and $0_q$ be the $q\times q$ identity and $q\times 1$ zero vector, $\beta_e^{\mathrm{phy}}=\beta_e^{\mathrm{phy}}(x_e^{\mathrm{tx}})>0$ the true path gain with $\beta_e^{\mathrm{phy}}=\bar{\beta}$ when homogeneous, and $K_{\mathrm{R}}=K_{\mathrm{R}}(x_e^{\mathrm{tx}})\ge 0$ the Rician factor. At fixed $x_e^{\mathrm{tx}}\in\mathcal{X}$, let $\tilde{z}_e=\tilde{z}_e(x_e^{\mathrm{tx}})\sim\mathcal{CN}(0_M,I_M)$ be diffuse multipath and $R_{\mathrm{rx},e}=R_{\mathrm{rx}}(x_e^{\mathrm{tx}})\succeq 0$ the Hermitian correlation matrix with $\mathrm{tr}(R_{\mathrm{rx},e})=M$ and root $R_{\mathrm{rx},e}^{1/2}$. Under isotropic scattering, $R_{\mathrm{rx},e}=I_M$; under a first-order directional-cosine Gaussian angle-of-arrival spread $\sigma_\theta\ge 0$ in radians, entry $(m_{\mathrm{a}},n_{\mathrm{a}})\in\{0,\ldots,M-1\}^2$ with $\Delta_{mn}=m_{\mathrm{a}}-n_{\mathrm{a}}$ equals $e^{-\mathrm{j}2\pi\delta\Delta_{mn}u_e}$ times $\exp(-2[\pi\delta\sigma_\theta\Delta_{mn}\sqrt{1-u_e^2}]^2)$. The physical channel $h_e^{\mathrm{phy}}=h_e^{\mathrm{phy}}(x_e^{\mathrm{tx}})\in\mathbb{C}^{M\times 1}$ is~\cite{zeng2019accessing}
\begin{equation}
h_e^{\mathrm{phy}}=\sqrt{\frac{\beta_e^{\mathrm{phy}}}{M N_0 W F_n}}\!\left(\!\sqrt{\frac{K_{\mathrm{R}}}{K_{\mathrm{R}}+1}}\,a_{\mathrm{st}}(u_e)+\frac{R_{\mathrm{rx},e}^{1/2}\tilde{z}_e}{\sqrt{K_{\mathrm{R}}+1}}\!\right)\!.
\label{eq:rician_phy}
\end{equation}
In~\eqref{eq:rician_phy}, $K_{\mathrm{R}}=0$ with $R_{\mathrm{rx},e}=I_M$ yields Rayleigh non-line-of-sight (NLoS) fading~\cite{bjornson2015optimal} and $K_{\mathrm{R}}\to\infty$ with $\beta_e^{\mathrm{phy}}=\bar{\beta}$ recovers the pure line-of-sight (LoS) channel $h_e(x_e^{\mathrm{tx}})$, satisfying $\mathbb{E}[\|h_e^{\mathrm{phy}}(x_e^{\mathrm{tx}})\|^2\mid\beta_e^{\mathrm{phy}}]=\beta_e^{\mathrm{phy}}(x_e^{\mathrm{tx}})/(N_0 W F_n)$ for all $K_{\mathrm{R}}\ge 0$. Since diffuse realizations in $\tilde{z}_e(x_e^{\mathrm{tx}})$ and residual blockage perturbations are unknown before robot arrival~\cite{zeng2019accessing}, offline planning optimizes over $h_e(x_e^{\mathrm{tx}})$ with multi-scenario perturbations and a service reserve, whereas online reception and out-of-sample tests operate on $h_e^{\mathrm{phy}}(x_e^{\mathrm{tx}})$ plus pilot errors.

To capture gain-scaled imperfect CSI~\cite{bjornson2015optimal} with spatial correlation around $h_e(x_e^{\mathrm{tx}})$ during planning, let $\omega\in\Omega_{\mathrm{tr}}\cup\{0\}$ index scenarios, where $\omega=0$ is the nominal channel and $\Omega_{\mathrm{tr}}$ is a finite set of spatially correlated realizations with normalized MSE $\epsilon_{\mathrm{CSI}}\ge 0$. Each $\omega\in\Omega_{\mathrm{tr}}$ synthesizes $z^\omega(x)=\sum_{r=1}^{R_f}\xi_r^\omega\phi_r^\omega(x)\in\mathbb{C}^{M\times 1}$ from $R_f\ge 1$ random Fourier features $\phi_r^\omega(x)=\exp(\mathrm{j}((k_r^\omega)^\top x+\varphi_r^\omega))$ with correlation length $L_{\mathrm{corr}}>0$, $k_r^\omega\sim\mathcal{N}(0_2,L_{\mathrm{corr}}^{-2}I_2)$, $\varphi_r^\omega\sim\mathrm{Unif}[0,2\pi)$, and $\xi_r^\omega\sim\mathcal{CN}(0_M,I_M/R_f)$ so that $\mathbb{E}[z^\omega(x)z^\omega(x)^H]=I_M$, while $z^0(x)\equiv 0_M$. Scenario $\omega$'s channel estimate is
\begin{equation}
\hat{h}_e^\omega=h_e(x_e^{\mathrm{tx}})+\sqrt{\|h_e(x_e^{\mathrm{tx}})\|^2\epsilon_{\mathrm{CSI}}/M}\,z^\omega(x_e^{\mathrm{tx}}).
\label{eq:csi_error}
\end{equation}

Let $\mathcal{A}(t)=\{e\in\mathcal{E}: s_e\le t < f_e\}$ be the active set, and $(\tilde{h}_j,\hat{h}_j)$ the true--estimated pair: $(h_j,\hat{h}_j^\omega)$ in planning scenario $\omega$, and $(h_j^{\mathrm{phy}},\allowbreak h_j^{\mathrm{phy}}+\allowbreak\|h_j^{\mathrm{phy}}\|\sqrt{\epsilon_{\mathrm{CSI}}/M}\,z_j^{\mathrm{test}})$ under independent error $z_j^{\mathrm{test}}$ in evaluation. With LMMSE combiner $w_e=\Sigma_{\mathcal{A}(t)}^{-1}\hat{h}_e$ for $\Sigma_{\mathcal{A}}=I_M+\sum_{j\in\mathcal{A}}p_j\hat{h}_j\hat{h}_j^H$~\cite{bjornson2015optimal,tse1999linear}, SINR is
\begin{equation}
\gamma_e=\frac{p_e|w_e^H\tilde{h}_e|^2}{\|w_e\|^2+\sum_{j\in\mathcal{A}(t)\setminus\{e\}}p_j|w_e^H\tilde{h}_j|^2},
\label{eq:sinr}
\end{equation}
where $\gamma_e=0$ if $p_e=0$ or $w_e=0$, giving rate $R_e(\mathcal{A}(t))=\chi W\log_2(1+\gamma_e)$ with factor $\chi\in(0,1]$. At $\omega=0$ where $\hat{h}_j=\tilde{h}_j=h_j$, $\gamma_e^0=p_e h_e^H C_{-e}^{-1}h_e$ with $C_{-e}=I_M+\sum_{j\in\mathcal{A}(t)\setminus\{e\}}p_j h_j h_j^H$.

Sorting $\{s_e,f_e\}_{e\in\mathcal{E}}$ partitions $[0,H]$ into elementary intervals $\ell$ of duration $\Delta_\ell\ge 0$ with invariant $\mathcal{A}_\ell$ and $R_e(\mathcal{A}_\ell)$. Packet delivery requires integrated bit service
\begin{equation}
S_e=\sum_{\ell:\,e\in\mathcal{A}_\ell}\Delta_\ell R_e(\mathcal{A}_\ell)\ge D_e.
\label{eq:service}
\end{equation}
Let $S_e^\omega$ denote~\eqref{eq:service} evaluated under $\{\hat{h}_j^\omega\}$ in~\eqref{eq:csi_error}. With service reserve factor $\zeta\in(0,1)$, planning enforces
\begin{equation}
\zeta S_e^\omega\ge b_{i_e}+\rho_e D_{i_e},\;\forall e\in\mathcal{E},\;\forall\omega\in\Omega_{\mathrm{tr}}\cup\{0\}.
\label{eq:reserve}
\end{equation}

\subsection{Scheduling and Energy Consumption Model}
For task $e\in\mathcal{E}$, let $g_e\ge 0$ be its capture timestamp at $x_e^{\mathrm{s}}$, $t_i^{\mathrm{rec}}>0$ the BS reconstruction delay, and $d_e=f_e+t_{i_e}^{\mathrm{rec}}$ the BS completion timestamp. Let $e^-$ denote task $e$'s predecessor on $\pi_{k_e}$, with depot $x_k^0\in\mathcal{X}$, $x_{e^-}^{\mathrm{tx}}\triangleq x_{k_e}^0$, and $f_{e^-}\triangleq 0$ for $e=e_{k_e,1}$, and let $v_k>0$ be robot $k$'s maximum speed. Robot $k_e$ moves from $x_{e^-}^{\mathrm{tx}}$ to $x_e^{\mathrm{s}}$ before $g_e$, then compresses over $T_e^{\mathrm{cpu}}$ in~\eqref{eq:cpu_energy} while moving to $x_e^{\mathrm{tx}}$ before $s_e$~\cite{mei2006deployment,zeng2019accessing}. This yields
\begin{align}
g_e&\ge f_{e^-}+\|x_e^{\mathrm{s}}-x_{e^-}^{\mathrm{tx}}\|/v_{k_e},\;\; g_e\ge 0,\;\; f_e\ge s_e,\;\; d_e\le H,\label{eq:causal_move}\\
s_e&\ge g_e+T_e^{\mathrm{cpu}},\qquad s_e\ge g_e+\|x_e^{\mathrm{tx}}-x_e^{\mathrm{s}}\|/v_{k_e}.\label{eq:causal_deadline}
\end{align}
Let $L_k(\mathbf{x})=\sum_{e\in\pi_k}(\|x_e^{\mathrm{s}}-x_{e^-}^{\mathrm{tx}}\|+\allowbreak\|x_e^{\mathrm{tx}}-x_e^{\mathrm{s}}\|)$ be robot $k$'s tour length, $c_k^{\mathrm{mov}}>0$ its propulsion coefficient in $\text{J/m}$, $\eta_k^{\mathrm{PA}}\in(0,1]$ its power-amplifier (PA) efficiency, $p_k^{\mathrm{cir}}\ge 0$ its active radio-frequency (RF) circuit power, and $B_k>0$ its battery budget. With $E_k^{\mathrm{mov}}=c_k^{\mathrm{mov}}L_k(\mathbf{x})$ and $E_k^{\mathrm{cpu}}=\sum_{e\in\pi_k}E_{\mathrm{pkt},e}^{\mathrm{cpu}}$ from~\eqref{eq:cpu_energy}, per-robot energy consumption satisfies~\cite{mei2006deployment,mao2017survey,you2017energy,bjornson2015optimal}
\begin{equation}
E_k=E_k^{\mathrm{mov}}+E_k^{\mathrm{cpu}}+\underbrace{\textstyle\sum_{e\in\pi_k}(p_k^{\mathrm{cir}}+p_e/\eta_k^{\mathrm{PA}})(f_e-s_e)}_{E_k^{\mathrm{rad}}}\le B_k.
\label{eq:energy}
\end{equation}

\subsection{Freshness Utility and Problem Formulation}
\label{subsec:utility}
Let $q_{0,i}\in(0,1]$ be reference sensing quality, $d_{0,i}>0$ the half-sensitivity distance, and $\psi_i:[\rho_{\min,i},1]\to(0,1]$ a continuously differentiable, non-decreasing, concave fidelity function with $\psi_i(1)=1$. We adopt $\psi_i(\rho)=(1-2^{-\alpha_i\rho})/(1-2^{-\alpha_i})$ with $\alpha_i>0$, recovering $\psi_i\equiv 1$ as $\alpha_i\to\infty$~\cite{zhao2025energy,zhu2025scene}. Task $e$'s planned quality is $q_e(\rho_e,x_e^{\mathrm{s}})=\psi_{i_e}(\rho_e)q_e^{\mathrm{sp}}(x_e^{\mathrm{s}})$, with spatial factor
\begin{equation}
q_e^{\mathrm{sp}}(x_e^{\mathrm{s}})=\frac{q_{0,i_e}}{1+\|x_e^{\mathrm{s}}-\tilde{y}_{i_e}\|^2/d_{0,i_e}^2}.
\label{eq:quality}
\end{equation}
Evaluation uses realized $y_{i_e}^{(w)}$ in~\eqref{eq:sensing_area} and~\eqref{eq:quality}, admitting packets satisfying~\eqref{eq:sensing_area} and~\eqref{eq:service}. With freshness constant $\tau_i>0$, let $e_i(t)$ denote the valid packet with latest $g_e$ among those reconstructed by time $t$, i.e., $d_e\le t$ with ties broken by larger $e$. Instantaneous target value decays with age $t-g_{e_i(t)}$~\cite{yates2021age,kountouris2021semantics}:
\begin{equation}
Q_i(t)=\begin{cases}
q_{e_i(t)}\,e^{-(t-g_{e_i(t)})/\tau_i}, & \text{if }e_i(t)\text{ exists},\\
0, & \text{otherwise}.
\end{cases}
\label{eq:inst_value}
\end{equation}
Ordering target $i$'s valid updates by $(d_e,e)$ and keeping those advancing $(g_e,e)$ gives $\mathcal{D}_i\subseteq\{e\in\mathcal{E}:i_e=i\}$, where $e\in\mathcal{D}_i$ holds over $[d_e,t_e^{\mathrm{next}})$ and $t_e^{\mathrm{next}}=H$ for the last valid update. Integrating~\eqref{eq:inst_value} over $[0,H]$ yields
\begin{equation}
U_i=\sum_{e\in\mathcal{D}_i}\frac{\psi_{i_e}(\rho_e)\,q_e^{\mathrm{sp}}(x_e^{\mathrm{s}})\,\tau_i}{H}\Big(e^{(g_e-d_e)/\tau_i}-e^{(g_e-t_e^{\mathrm{next}})/\tau_i}\Big).
\label{eq:utility}
\end{equation}
To ensure target-level fairness, we maximize $U=\min_{i\in\mathcal{I}} U_i$, linked to Bayesian risk by Lemma~\ref{lem:markov_risk}.

\begin{lemma}
\label{lem:markov_risk}
Suppose each target $i\in\mathcal{I}$ has a latent binary state $Z_i(t)\in\{0,1\}$ evolving over $[0,H]$ as a stationary symmetric continuous-time Markov chain with switching rate $\kappa_i^{\mathrm{st}}>0$. Assume timestamps and validity indicators are exogenous of $\{Z_i(t)\}$ and observation errors, and each valid packet $e\in\mathcal{D}_i$ yields a decoded estimate $\hat{Z}_e$ of $Z_i(g_e)$ with symmetric error probability $\varepsilon_e(\rho_e,x_e^{\mathrm{s}})\in[0,1/2]$, conditionally independent of $\{Z_i(t):t>g_e\}$ given $Z_i(g_e)$. Under the latest-valid-packet rule, let $\mathcal{I}_i(t)=(e_i(t),\hat{Z}_{e_i(t)},g_{e_i(t)},d_{e_i(t)})$ denote the BS information set, which includes the received packet identity $e_i(t)$ and its known error probability $\varepsilon_{e_i(t)}$, when $e_i(t)$ exists, and $\mathcal{I}_i(t)=\emptyset$ otherwise, with instantaneous minimum Bayes $0$--$1$ risk $\mathcal{R}_i(t)=\min_{\hat{z}\in\{0,1\}}\mathbb{P}(Z_i(t)\ne\hat{z}\mid\mathcal{I}_i(t))$. Setting $q_e(\rho_e,x_e^{\mathrm{s}})=1-2\varepsilon_e(\rho_e,x_e^{\mathrm{s}})$ and $\tau_i=1/(2\kappa_i^{\mathrm{st}})$ yields $\mathcal{R}_i(t)=(1-Q_i(t))/2$ and
\begin{equation}
\bar{\mathcal{R}}_i\triangleq\frac{1}{H}\int_0^H\mathcal{R}_i(t)\,\mathrm{d}t=\frac{1-U_i}{2}.
\label{eq:bayes_risk}
\end{equation}
Hence maximizing $\min_{i\in\mathcal{I}}U_i$ is equivalent to minimizing worst-target Bayesian risk $\max_{i\in\mathcal{I}}\bar{\mathcal{R}}_i$.
\end{lemma}
\begin{IEEEproof}
Before first delivery, the uniform prior gives $\mathcal{R}_i(t)=1/2=(1-Q_i(t))/2$. On $[d_e,t_e^{\mathrm{next}})$ with $e=e_i(t)\in\mathcal{D}_i$, the symmetric Markov kernel gives $\mathbb{P}(Z_i(t)=Z_i(g_e)\mid Z_i(g_e))=(1+e^{-2\kappa_i^{\mathrm{st}}(t-g_e)})/2$. By symmetry, exogenous validity, and conditional independence, the maximum a posteriori (MAP) reliability of $\hat{Z}_e$ is $(1-\varepsilon_e)(1+e^{-2\kappa_i^{\mathrm{st}}(t-g_e)})/2+\varepsilon_e(1-e^{-2\kappa_i^{\mathrm{st}}(t-g_e)})/2=[1+(1-2\varepsilon_e)e^{-2\kappa_i^{\mathrm{st}}(t-g_e)}]/2$. Substituting $q_e=1-2\varepsilon_e$ and $2\kappa_i^{\mathrm{st}}=1/\tau_i$ gives $\mathcal{R}_i(t)=(1-Q_i(t))/2$, and integration over $[0,H]$ yields~\eqref{eq:bayes_risk}.
\end{IEEEproof}

Collecting waypoints $\mathbf{x}=\{x_e\}_{e\in\mathcal{E}}$, compression ratios $\boldsymbol{\rho}=\{\rho_e\}_{e\in\mathcal{E}}$, powers $\mathbf{p}=\{p_e\}_{e\in\mathcal{E}}$, timestamps $\mathbf{t}=(\mathbf{g},\mathbf{s},\mathbf{f})$ with $\mathbf{g}=\{g_e\}_{e\in\mathcal{E}}$, $\mathbf{s}=\{s_e\}_{e\in\mathcal{E}}$, and $\mathbf{f}=\{f_e\}_{e\in\mathcal{E}}$, and hypograph variable $u\ge 0$, we formulate
\begin{subequations}
\label{eq:problem}
\begin{align}
(\mathcal{P})\;\;&\max_{\Pi,\mathbf{x},\boldsymbol{\rho},\mathbf{p},\mathbf{t},u}\;u\label{eq:prob_a}\\
\text{s.t.}\;\;&u\le U_i(\Pi,\mathbf{x},\boldsymbol{\rho},\mathbf{p},\mathbf{t}),\quad \forall i\in\mathcal{I},\label{eq:prob_b}\\
&d_{\min,i_e}+r_y\le\|x_e^{\mathrm{s}}-\tilde{y}_{i_e}\|\le d_{\max,i_e}-r_y,\;\forall e\in\mathcal{E},\label{eq:prob_c}\\
&x_e^{\mathrm{s}},x_e^{\mathrm{tx}}\in\mathcal{X},\;\rho_{\min,i_e}\le\rho_e\le 1,\; 0\le p_e\le p_{\max,k_e},\label{eq:prob_d}\\
&\zeta S_e^\omega\ge b_{i_e}+\rho_e D_{i_e},\;\forall e\in\mathcal{E},\;\forall\omega\in\Omega_{\mathrm{tr}}\cup\{0\},\label{eq:prob_e}\\
&E_k^{\mathrm{mov}}+E_k^{\mathrm{cpu}}+E_k^{\mathrm{rad}}\le B_k,\;\forall k\in\mathcal{K},\label{eq:prob_bat}\\
&\eqref{eq:causal_move}\text{--}\eqref{eq:causal_deadline},\;\{\pi_k\}_{k\in\mathcal{K}}\text{ partitions }\mathcal{E},\;\bigcup_{e\in\mathcal{E}}\{i_e\}=\mathcal{I},\;u\ge 0,\label{eq:prob_f}
\end{align}
\end{subequations}
where $\mathcal{F}_{\mathrm{tr}}$ denotes~\eqref{eq:prob_c}--\eqref{eq:prob_f} and $x_e^{\mathrm{tx}}=x_e^{\mathrm{s}}$ defines $\mathcal{F}_{\mathrm{tr}}^{\mathrm{col}}\subset\mathcal{F}_{\mathrm{tr}}$, maximizing~\eqref{eq:prob_a}--\eqref{eq:prob_b} under bounds~\eqref{eq:prob_c}--\eqref{eq:prob_d}, service~\eqref{eq:prob_e}, battery~\eqref{eq:prob_bat} via~\eqref{eq:energy}, and causality~\eqref{eq:prob_f}. Problem~\eqref{eq:problem} is a non-convex MINLP: even at fixed $\Pi$, shifting $\mathbf{t}$ alters $\mathcal{A}_\ell$ in~\eqref{eq:sinr}--\eqref{eq:service},~\eqref{eq:utility} couples $(\boldsymbol{\rho},\mathbf{x}^{\mathrm{s}},\mathbf{t})$ non-convexly, and~\eqref{eq:workload} changes slope at breakpoints.

\section{Algorithm Design}
\label{sec:algorithm}

We construct $\widehat{\Pi}$ in Section~\ref{subsec:init} and solve~$(\mathcal{P}[\widehat{\Pi}])$ via a three-block BCD--SCA algorithm~\cite{zhao2025energy,razaviyayn2013unified,sun2017majorization}. At iteration $m\in\{0,\ldots,L_{\max}-1\}$ with $L_{\max}\ge 1$, Block~A optimizes compression and scheduling on a fixed branch, Block~B refines powers and adjacent ordering via dual sensitivities, and Block~C updates waypoints via SOCP~\cite{lobo1998applications}. In Blocks~A and~B, $\mathbf{x}^{(m)}=\{(x_e^{\mathrm{s},(m)},x_e^{\mathrm{tx},(m)})\}_{e\in\mathcal{E}}$ fixes $L_e^{\mathrm{pre},(m)}=\|x_e^{\mathrm{s},(m)}-x_{e^-}^{\mathrm{tx},(m)}\|$, $L_e^{\mathrm{tr},(m)}=\|x_e^{\mathrm{tx},(m)}-x_e^{\mathrm{s},(m)}\|$, $q_e^{\mathrm{sp},(m)}=q_e^{\mathrm{sp}}(x_e^{\mathrm{s},(m)})$, $\{\hat{h}_e^{\omega,(m)}\}$, and $\hat{B}_k^{(m)}=B_k-c_k^{\mathrm{mov}}\sum_{e\in\pi_k}(L_e^{\mathrm{pre},(m)}+L_e^{\mathrm{tr},(m)})-\sum_{e\in\pi_k}E_{0,i_e}$.

\subsection{Semantic Compression and Scheduling Optimization}
\label{subsec:blockA}
With fixed $\mathbf{p}^b=\mathbf{p}^{(m)}$ and $\mathbf{x}^{(m)}$, let $\mathcal{E}_i=\{e\in\mathcal{E}:i_e=i\}$ and $(\boldsymbol{\rho}^b,\mathbf{t}^b)$ be a feasible anchor initialized at $(\boldsymbol{\rho}^{(m)},\mathbf{t}^{(m)})$. An event branch $\beta$ from $\mathbf{t}^b$ fixes three linear orderings. First, $(t_{\beta_1},\ldots,t_{\beta_{2P}})$ orders $\{s_e,f_e\}_{e\in\mathcal{E}}$ into intervals $\ell\in\{1,\ldots,2P-1\}$ of duration $\Delta_\ell^\beta=t_{\beta_{\ell+1}}-t_{\beta_\ell}\ge 0$ with constant rates $R_{e\ell}^{\omega,b}$. Second, $\prec_g$ orders captures on $\mathcal{E}_i$ with margin $\epsilon_{i,r}^g>0$ when reversing task indices. Third, $\prec_d$ orders completions on $\mathcal{E}_i$ under canonical tie-breaking, keeping $U_i$ continuous since superseded updates have zero active duration. Fixing $(\prec_g,\prec_d)$ keeps $\mathcal{D}_i^\beta$ invariant and $t_e^{\mathrm{next}}(\mathbf{t})$ affine on $\beta$, making~\eqref{eq:reserve} affine in $(\mathbf{t},\boldsymbol{\rho})$:
\begin{equation}
\zeta S_e^{\omega,\beta}(\mathbf{t};\mathbf{p}^b)\ge b_{i_e}+\rho_e D_{i_e},\quad \forall e\in\mathcal{E},\;\forall\omega\in\Omega_{\mathrm{tr}}\cup\{0\}.
\label{eq:affine_service}
\end{equation}
With active RF power draw $\kappa_e^b=p_{k_e}^{\mathrm{cir}}+p_e^b/\eta_{k_e}^{\mathrm{PA}}$ and epigraph variables $\mathbf{c}=\{c_e\}_{e\in\mathcal{E}}$ in~\eqref{eq:epigraph}, causal precedence, transit, and battery constraints become
\begin{align}
g_e&\ge f_{e^-}+L_e^{\mathrm{pre},(m)}/v_{k_e},\;\; s_e\ge g_e+t_{0,i_e}+c_e/F_{k_e},\label{eq:lin_cpu}\\
s_e&\ge g_e+L_e^{\mathrm{tr},(m)}/v_{k_e},\label{eq:lin_tr}\\
\sum_{e\in\pi_k}&\big(\epsilon_k c_e+\kappa_e^b(f_e-s_e)\big)\le \hat{B}_k^{(m)},\quad \forall k\in\mathcal{K}.\label{eq:lin_battery}
\end{align}
Collecting~\eqref{eq:payload},~\eqref{eq:epigraph},~\eqref{eq:affine_service}--\eqref{eq:lin_battery},~\eqref{eq:causal_move}--\eqref{eq:causal_deadline}, and the ordering inequalities of $\beta$ defines the closed polyhedron $\mathcal{C}_\beta(\mathbf{x}^{(m)},\mathbf{p}^b)$.

To convexify the difference-of-exponentials utility~\eqref{eq:utility} around $(\boldsymbol{\rho}^b,\mathbf{t}^b)$ with $c_e^b=\Gamma_{i_e}(\rho_e^b)$, define $a_e^b=\psi_{i_e}(\rho_e^b)\allowbreak q_e^{\mathrm{sp},(m)}\allowbreak\tau_{i_e}/H$, $A_e(\mathbf{t})=(g_e-d_e)/\tau_{i_e}$, and $B_e(\mathbf{t})=(g_e-t_e^{\mathrm{next}}(\mathbf{t}))/\tau_{i_e}$. First-order Taylor linearization of $e^{A_e(\mathbf{t})}$ at $A_e^b=A_e(\mathbf{t}^b)$ while retaining $-e^{B_e(\mathbf{t})}$ yields the concave scheduling minorant~\cite{sun2017majorization}
\begin{equation}
\hat{U}_i(\mathbf{t};\boldsymbol{\rho}^b,\mathbf{t}^b)=\sum_{e\in\mathcal{D}_i^\beta}a_e^b\big\{e^{A_e^b}[1+A_e(\mathbf{t})-A_e^b]-e^{B_e(\mathbf{t})}\big\}.
\label{eq:minorant}
\end{equation}

\begin{lemma}
\label{lem:minorant}
For fixed $\boldsymbol{\rho}^b$, $\hat{U}_i(\mathbf{t};\boldsymbol{\rho}^b,\mathbf{t}^b)$ in~\eqref{eq:minorant} is concave in $\mathbf{t}$ and minorizes $U_i^\beta(\boldsymbol{\rho}^b,\mathbf{t})$ globally with matching value and gradient at $\mathbf{t}^b$.
\end{lemma}
\begin{IEEEproof}
Convexity of $e^x$ gives $e^{A_e(\mathbf{t})}\ge e^{A_e^b}(1+A_e(\mathbf{t})-A_e^b)$ with equality and matching gradient at $\mathbf{t}^b$; negativity of the Hessian $\nabla_{\mathbf{t}}^2\hat{U}_i=-\sum_{e\in\mathcal{D}_i^\beta}a_e^b e^{B_e(\mathbf{t})}(\nabla_{\mathbf{t}} B_e)(\nabla_{\mathbf{t}} B_e)^\top\preceq 0$ proves concavity.
\end{IEEEproof}

Let $\widehat{E}_{\mathrm{tot}}^{(m)}(\mathbf{c},\mathbf{t})=\sum_{k\in\mathcal{K}}[c_k^{\mathrm{mov}}L_k(\mathbf{x}^{(m)})+\sum_{e\in\pi_k}(E_{0,i_e}+\epsilon_k c_e+\kappa_e^b(f_e-s_e))]$ denote the lifted total swarm energy, $\eta_{\mathrm{sec}},\eta_E>0$ small tie-breaking weights for mean utility and energy, $\eta_\rho,\eta_t>0$ proximal regularization weights, and $\varepsilon_{\mathrm{fl}}\ge 0$ a floor relaxation tolerance defining $u_{\mathrm{fl}}^b=\max\{0,U^\beta(\boldsymbol{\rho}^b,\mathbf{t}^b)-\varepsilon_{\mathrm{fl}}\}$; implementation sets $\varepsilon_{\mathrm{fl}}=10^{-6}$ with exact acceptance $U^{(m+1)}\ge U^{(m)}$ enforced after solving and $u_{\mathrm{fl}}^b=0$ during Block-B power trials. Block~A executes one of two exact convex paths according to $\psi_i$:

1)~\emph{Quality-preserving and exponential-saturation regimes:} When $\psi_i\equiv 1$, $\hat{U}_i(\mathbf{t};\boldsymbol{\rho}^b,\mathbf{t}^b)$ is independent of $\boldsymbol{\rho}$ and written $\hat{U}_i(\mathbf{t};\mathbf{t}^b)$. For $\psi_i(\rho)=(1-2^{-\alpha_i\rho})/(1-2^{-\alpha_i})$ with $\tilde{\alpha}_i=\alpha_i\ln 2$, expanding $(1-e^{-\tilde{\alpha}_i\rho_e})(e^{A_e}-e^{B_e})=(e^{A_e}+e^{B_e-\tilde{\alpha}_i\rho_e})-(e^{B_e}+e^{A_e-\tilde{\alpha}_i\rho_e})$ yields a joint difference-of-exponentials concave minorant $\hat{U}_i^{\mathrm{dc}}(\boldsymbol{\rho},\mathbf{t};\boldsymbol{\rho}^b,\mathbf{t}^b)\le U_i^\beta(\boldsymbol{\rho},\mathbf{t})$ tight at $(\boldsymbol{\rho}^b,\mathbf{t}^b)$. Block~A solves one exponential-cone program, replacing $\hat{U}_i$ with $\hat{U}_i^{\mathrm{dc}}$ for finite $\alpha_i$:
\begin{equation}
\begin{aligned}
\max_{\boldsymbol{\rho},\mathbf{c},\mathbf{t},u}\;\; &u+\tfrac{\eta_{\mathrm{sec}}}{N}\!\sum_{i\in\mathcal{I}}\hat{U}_i(\mathbf{t};\mathbf{t}^b)-\eta_E\widehat{E}_{\mathrm{tot}}^{(m)}(\mathbf{c},\mathbf{t})\\[-2pt]
&\quad-\tfrac{\eta_\rho}{2}\|\boldsymbol{\rho}-\boldsymbol{\rho}^b\|^2-\tfrac{\eta_t}{2}\|\mathbf{t}-\mathbf{t}^b\|^2\\[-1pt]
\text{s.t.}\;\; &u_{\mathrm{fl}}^b\le u\le\hat{U}_i(\mathbf{t};\mathbf{t}^b),\;\forall i\in\mathcal{I},\\[-1pt]
&(\boldsymbol{\rho},\mathbf{c},\mathbf{t})\in\mathcal{C}_\beta(\mathbf{x}^{(m)},\mathbf{p}^b),
\end{aligned}
\label{eq:sub_Qt}
\end{equation}
followed by exact elimination $c_e^{(m+1)}=\Gamma_{i_e}(\rho_e^{(m+1)})$ and capture tightening $g_e^{(m+1)}\leftarrow\max\{g_e,\min\{s_e-t_{0,i_e}-c_e^{(m+1)}/F_{k_e},s_e-L_e^{\mathrm{tr},(m)}/v_{k_e}\}\}$ subject to $\prec_g$ on $\mathcal{E}_i$.

2)~\emph{General concave fidelity with conic hypograph:} Block~A first solves~\eqref{eq:sub_Qt} over $(\mathbf{t},u)$ at fixed $(\boldsymbol{\rho}^b,\mathbf{c}^b)$ to obtain $\tilde{\mathbf{t}}$, then fixes $\tilde{\mathbf{t}}$ with non-negative weights $\varpi_e(\tilde{\mathbf{t}})=q_e^{\mathrm{sp},(m)}\tau_{i_e}(e^{A_e(\tilde{\mathbf{t}})}-e^{B_e(\tilde{\mathbf{t}})})/H$ for $e\in\mathcal{D}_i^\beta$ and $0$ otherwise, sets floor $\tilde{u}_{\mathrm{fl}}=\max\{0,U^\beta(\boldsymbol{\rho}^b,\tilde{\mathbf{t}})-\varepsilon_{\mathrm{fl}}\}$, and solves
\begin{equation}
\begin{aligned}
\max_{\boldsymbol{\rho},\mathbf{c},u}\quad &u-\tfrac{\eta_\rho}{2}\|\boldsymbol{\rho}-\boldsymbol{\rho}^b\|^2\\
\text{s.t.}\quad &\tilde{u}_{\mathrm{fl}}\le u\le\sum_{e\in\mathcal{D}_i^\beta}\varpi_e(\tilde{\mathbf{t}})\,\psi_{i_e}(\rho_e),\quad \forall i\in\mathcal{I},\\
&(\boldsymbol{\rho},\mathbf{c},\tilde{\mathbf{t}})\in\mathcal{C}_\beta(\mathbf{x}^{(m)},\mathbf{p}^b),
\end{aligned}
\label{eq:sub_Q}
\end{equation}
and refreshes $(\mathbf{t},u)$ via~\eqref{eq:sub_Qt} at $(\boldsymbol{\rho}^{(m+1)},\mathbf{c}^{(m+1)})$.

Assume Slater's condition holds on~\eqref{eq:sub_Qt}--\eqref{eq:sub_Q}; if Block~A fails to yield an accepted iterate with finite downstream duals $(\boldsymbol{\mu}^\star,\boldsymbol{\nu}^\star,\boldsymbol{\theta}^{\mathrm{pre},\star},\boldsymbol{\theta}^\star,\boldsymbol{\theta}^{\mathrm{tr},\star},\boldsymbol{\mu}^{\mathrm{ord},\star})$, prior duals are reused when available and Block~B is skipped otherwise. Let $\lambda_i^\star$, $\mu_{e,\omega}^\star$, $\xi_{e,a}^\star$, $(\theta_e^{\mathrm{pre},\star},\theta_e^\star,\theta_e^{\mathrm{tr},\star})$, $\nu_k^\star$, and $\mu_\ell^{\mathrm{ord},\star}\ge 0$ be finite optimal duals of the utility hypograph,~\eqref{eq:affine_service},~\eqref{eq:epigraph},~\eqref{eq:lin_cpu}--\eqref{eq:lin_tr},~\eqref{eq:lin_battery}, and $\Delta_\ell^\beta\ge 0$, with $\tilde{\lambda}_i^\star=\lambda_i^\star+\eta_{\mathrm{sec}}/N$ and $\tilde{\nu}_k^\star=\nu_k^\star+\eta_E$ in~\eqref{eq:sub_Qt}, or $\lambda_i^\star$ and $\nu_k^\star$ in~\eqref{eq:sub_Q}. Let $F_e^{\mathrm{fid}}(r)$ be packet $e$'s concave fidelity minorant with derivative $M_e^{\mathrm{fid}}(r)=(F_e^{\mathrm{fid}})'(r)$, equal to $0$ if $\psi_i\equiv 1$ and $\varpi_e(\mathbf{t}^\star)\psi_{i_e}'(\rho_e^\star)$ at a stationary anchor. Isolating $(\rho_e,c_e)$ in minimization form gives
\begin{align}
\mathcal{L}_{\rho,c}=\sum_{e\in\mathcal{E}}\Big\{&-\tilde{\lambda}_{i_e}^\star F_e^{\mathrm{fid}}(\rho_e)+\!\sum_\omega\mu_{e,\omega}^\star(b_{i_e}+\rho_e D_{i_e}-\zeta S_e^{\omega,\beta})\notag\\
&+\!\sum_{a=1}^A\xi_{e,a}^\star(A_{i_e a}\rho_e+B_{i_e a}-c_e)+\tfrac{\eta_\rho}{2}(\rho_e-\rho_e^b)^2\notag\\
&+\big(\epsilon_{k_e}\tilde{\nu}_{k_e}^\star+\theta_e^\star/F_{k_e}\big)c_e\Big\}.
\label{eq:part_lag}
\end{align}
Define wireless price $\Phi_e^{\mathrm{com}}=D_{i_e}\sum_\omega\mu_{e,\omega}^\star\ge 0$, CPU price $\Psi_e^{\mathrm{cpu}}=\epsilon_{k_e}\tilde{\nu}_{k_e}^\star+\theta_e^\star/F_{k_e}\ge 0$, regularized net price $\widetilde{\Phi}_e^{\mathrm{net}}(r)=\Phi_e^{\mathrm{com}}-\tilde{\lambda}_{i_e}^\star M_e^{\mathrm{fid}}(r)$, and unregularized net price $\widetilde{\Phi}_e^{0,\mathrm{net}}(r)=\Phi_e^{\mathrm{com}}-\tilde{\lambda}_{i_e}^\star\varpi_e\psi_{i_e}'(r)$. Writing $\rho_{\min}=\rho_{\min,i_e}$ and denoting the normal cone to $[\rho_{\min},1]$ by $N_{[\rho_{\min},1]}$, KKT optimality of~\eqref{eq:part_lag} requires
\begin{equation}
0\in\widetilde{\Phi}_e^{\mathrm{net}}(\rho_e^\star)+\eta_\rho(\rho_e^\star-\rho_e^b)+\Psi_e^{\mathrm{cpu}}\partial\Gamma_{i_e}(\rho_e^\star)+N_{[\rho_{\min},1]}(\rho_e^\star).
\label{eq:kkt_inc}
\end{equation}

For $A=3$ in~\eqref{eq:workload} on pieces $I_{\varsigma_3}=[\rho_{\min},\varrho_2]$, $I_{\varsigma_2}=[\varrho_2,\varrho_1]$, and $I_{\varsigma_1}=[\varrho_1,1]$ with $0<\varsigma_1<\varsigma_2<\varsigma_3$, define the non-decreasing ratio $\tilde{\vartheta}_e^0(r)=\widetilde{\Phi}_e^{0,\mathrm{net}}(r)/(c_{i_e}\Psi_e^{\mathrm{cpu}})$ for $\Psi_e^{\mathrm{cpu}}>0$ and root sets $\mathcal{S}_e^0(\varsigma,I)=\{r\in I:\tilde{\vartheta}_e^0(r)=\varsigma\}$, reducing to $\vartheta_e=\Phi_e^{\mathrm{com}}/(c_{i_e}\Psi_e^{\mathrm{cpu}})$ when $\psi_{i_e}'\equiv 0$. Let $J_e(r)=-\tilde{\lambda}_{i_e}^\star F_e^{\mathrm{fid}}(r)+\Phi_e^{\mathrm{com}}r+\Psi_e^{\mathrm{cpu}}\Gamma_{i_e}(r)+\frac{\eta_\rho}{2}(r-\rho_e^b)^2$, written $J_e^0$ at $\eta_\rho=0$ and $F_e^{\mathrm{fid}}=\varpi_e\psi_{i_e}$, and $\mathrm{Proj}_{\mathcal{S}}(\cdot)$ be Euclidean projection onto $\mathcal{S}$.

\begin{proposition}
\label{prop:kkt}
Assume finite optimal duals exist at stationary anchor $(\boldsymbol{\rho}^b,\mathbf{t}^b)=(\boldsymbol{\rho}^\star,\mathbf{t}^\star)$ or at $\eta_\rho=0$ in~\eqref{eq:sub_Q}. By Appendix~\ref{app:proof_prop1}, for $\Psi_e^{\mathrm{cpu}}>0$, $\rho_e^\star$ obeys:
\begin{equation}
\rho_e^\star\in\begin{cases}
\{1\}, & \tilde{\vartheta}_e^0(1)<\varsigma_1,\\
\mathcal{S}_e^0(\varsigma_1,I_{\varsigma_1}), & \tilde{\vartheta}_e^0(\varrho_1)\le \varsigma_1\le\tilde{\vartheta}_e^0(1),\\
\{\varrho_1\}, & \varsigma_1<\tilde{\vartheta}_e^0(\varrho_1)<\varsigma_2,\\
\mathcal{S}_e^0(\varsigma_2,I_{\varsigma_2}), & \tilde{\vartheta}_e^0(\varrho_2)\le \varsigma_2\le\tilde{\vartheta}_e^0(\varrho_1),\\
\{\varrho_2\}, & \varsigma_2<\tilde{\vartheta}_e^0(\varrho_2)<\varsigma_3,\\
\mathcal{S}_e^0(\varsigma_3,I_{\varsigma_3}), & \tilde{\vartheta}_e^0(\rho_{\min})\le \varsigma_3\le\tilde{\vartheta}_e^0(\varrho_2),\\
\{\rho_{\min}\}, & \tilde{\vartheta}_e^0(\rho_{\min})>\varsigma_3,
\end{cases}
\label{eq:prop1_regimes}
\end{equation}
while $\Psi_e^{\mathrm{cpu}}=0$ attains a boundary or zero-root of $\widetilde{\Phi}_e^{0,\mathrm{net}}$ on $[\rho_{\min},1]$. For $\eta_\rho>0$, $\rho_e^\star=\arg\min_{r\in\{r_{\varsigma_3},r_{\varsigma_2},r_{\varsigma_1}\}}J_e(r)$ over unique piece minimizers $r_\varsigma\in I_\varsigma$, closing under $M_e^{\mathrm{fid}}\equiv 0$ to
\begin{equation}
r_\varsigma=\mathrm{Proj}_{I_\varsigma}\!\left(\rho_e^b+\frac{\varsigma\,c_{i_e}\Psi_e^{\mathrm{cpu}}-\Phi_e^{\mathrm{com}}}{\eta_\rho}\right),\quad \varsigma\in\{\varsigma_3,\varsigma_2,\varsigma_1\}.
\label{eq:prop1_proj}
\end{equation}
Under strictly concave $\psi_i$, at breakpoint $\varrho_a$ for $a\in\{1,2\}$ with $\vartheta_e\in(\varsigma_a,\varsigma_{a+1})$ and $\tilde{\lambda}_e=\tilde{\lambda}_{i_e}^\star\varpi_e/(c_{i_e}\Psi_e^{\mathrm{cpu}})>0$, a stationary anchor satisfies $\rho_e^\star>\varrho_a$ if and only if $\psi_{i_e}'(\varrho_a)>(\vartheta_e-\varsigma_a)/\tilde{\lambda}_e$.
\end{proposition}

\begin{figure*}[t]
\centering
\includegraphics[width=.91\textwidth]{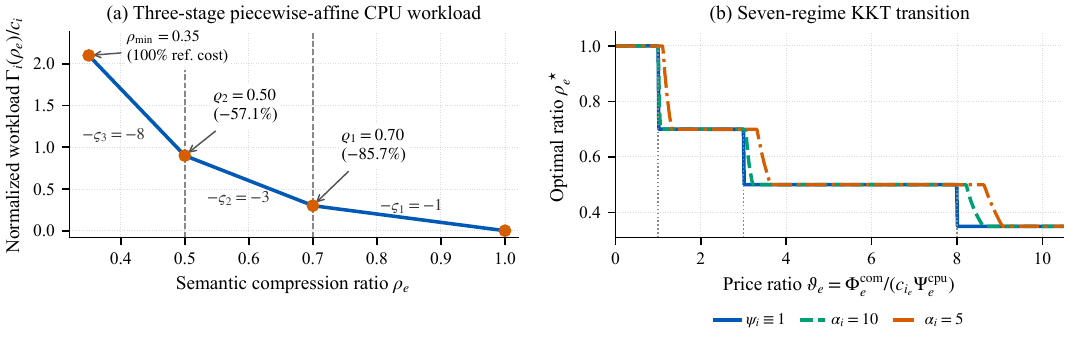}
\caption{CPU workload versus compression ratio and optimal compression ratio versus relative price under quality-preserving and lossy pruning.}
\label{fig:kkt_kink}
\end{figure*}

Fig.~\ref{fig:kkt_kink} illustrates Proposition~\ref{prop:kkt} for $\eta_\rho=0$, $(\varrho_1,\varrho_2,\rho_{\min})=(0.70,0.50,0.35)$, and $(\varsigma_1,\varsigma_2,\varsigma_3)=(1,3,8)$. Under quality-preserving pruning with $\psi_i\equiv 1$, $\rho_e^\star$ stays at a workload breakpoint across each open price interval $(\varsigma_a,\varsigma_{a+1})$, whereas steeper fidelity loss with $\alpha_i\in\{5,10\}$ and $\tilde{\lambda}_e=1$ shifts $\rho_e^\star$ above $\varrho_a$ wherever $\vartheta_e<\varsigma_a+\tilde{\lambda}_e\psi_i'(\varrho_a)$ in Proposition~\ref{prop:kkt}.

\begin{remark}
\label{rem:kink}
For the joint $(\boldsymbol{\rho},\mathbf{c},\mathbf{t},u)$ subproblem~\eqref{eq:sub_Qt}, $\rho_e$ and $f_e$ share one optimal dual set. Let $\mathcal{A}_e^{\mathrm{end}}$ denote the active set on the elementary interval immediately following $f_e$ in $\beta$'s endpoint sequence, with $\mathcal{A}_e^{\mathrm{end}}=\emptyset$ if $f_e$ is the last endpoint, ensuring well-defined one-sided branch derivatives even at coincident endpoints. Differentiating the Lagrangian of~\eqref{eq:sub_Qt} with respect to $f_e$ yields $\zeta\sum_\omega\mu_{e,\omega}^\star R_e^{\omega,b}(\mathcal{A}_e^{\mathrm{end}}\cup\{e\})=\tilde{\nu}_{k_e}^\star\kappa_e^b+\Xi_e^{\mathrm{int},\star}+\Theta_e^{\mathrm{sched},\star}$. Here, $\Xi_e^{\mathrm{int},\star}=\zeta\sum_{j\in\mathcal{A}_e^{\mathrm{end}},\omega}\mu_{j,\omega}^\star[R_j^{\omega,b}(\mathcal{A}_e^{\mathrm{end}})-R_j^{\omega,b}(\mathcal{A}_e^{\mathrm{end}}\cup\{e\})]$ measures the co-channel interference externality, which is non-negative at $\omega=0$ though potentially signed under mismatched CSI $\omega\in\Omega_{\mathrm{tr}}$, and $\Theta_e^{\mathrm{sched},\star}$ collects successor travel dual $+\theta_{e^+}^{\mathrm{pre},\star}$ for successor $e^+$ on $\pi_{k_e}$, branch-ordering, deadline $d_e\le H$, freshness, and proximal $\eta_t(f_e^\star-f_e^b)$ terms at $f_e^\star$. Whenever $\sum_\omega\mu_{e,\omega}^\star>0$, $\Psi_e^{\mathrm{cpu}}>0$, and $\bar{R}_e^\star=\sum_\omega\mu_{e,\omega}^\star R_e^{\omega,b}(\mathcal{A}_e^{\mathrm{end}}\cup\{e\})/\sum_\omega\mu_{e,\omega}^\star>0$, defining $\omega_e^{\mathrm{cpu},\star}=\epsilon_{k_e}\tilde{\nu}_{k_e}^\star/\Psi_e^{\mathrm{cpu}}\in(0,1]$ yields the wireless-to-CPU exchange law
\begin{equation}
\begin{aligned}
\vartheta_e=\frac{\Phi_e^{\mathrm{com}}}{c_{i_e}\Psi_e^{\mathrm{cpu}}}&=\omega_e^{\mathrm{cpu},\star}\frac{D_{i_e}\kappa_e^b}{c_{i_e}\epsilon_{k_e}\zeta\bar{R}_e^\star}\\
&\quad+\frac{D_{i_e}(\Xi_e^{\mathrm{int},\star}+\Theta_e^{\mathrm{sched},\star})}{c_{i_e}\Psi_e^{\mathrm{cpu}}\zeta\bar{R}_e^\star}.
\end{aligned}
\label{eq:exchange_law}
\end{equation}
In~\eqref{eq:exchange_law}, whenever intra-task transit exceeds CPU delay with $L_e^{\mathrm{tr},(m)}/v_{k_e}>T_e^{\mathrm{cpu}}$, complementary slackness forces $\theta_e^\star=0$ and $\Psi_e^{\mathrm{cpu}}=\epsilon_{k_e}\tilde{\nu}_{k_e}^\star$, so compression adds no schedule delay until computation matches transit time or reaches a steeper breakpoint in Fig.~\ref{fig:kkt_kink}.
\end{remark}

\begin{lemma}
\label{lem:lift_exact}
From any feasible anchor $(\boldsymbol{\rho}^b,\mathbf{t}^b)$ on branch $\beta$ with utility $U^b$, Block~A, with the anchor floor enforced and $\varepsilon_{\mathrm{fl}}=0$, produces an iterate $(\boldsymbol{\rho}^{(m+1)},\mathbf{c}^{(m+1)},\mathbf{t}^{(m+1)})\in\mathcal{C}_\beta(\mathbf{x}^{(m)},\mathbf{p}^b)$ with zero epigraph relaxation gap and $U(\boldsymbol{\rho}^{(m+1)},\mathbf{t}^{(m+1)})\ge U^b$.
\end{lemma}
\begin{IEEEproof}
By Lemma~\ref{lem:minorant} and $\hat{U}_i^{\mathrm{dc}}\le U_i^\beta$, global minorization and tightness at $(\boldsymbol{\rho}^b,\mathbf{t}^b)$ ensure anchor feasibility under floor $u\ge U^b$, so $U_i^\beta\ge u^\star\ge U^b$ holds across each substep of~\eqref{eq:sub_Qt}--\eqref{eq:sub_Q}. Setting $c_e^{(m+1)}=\Gamma_{i_e}(\rho_e^{(m+1)})$ and tightening $g_e$ preserve feasibility and $U_i^\beta$.
\end{IEEEproof}

\subsection{Transmit Power Optimization}
\label{subsec:blockB}
Block~B exploits the optimal dual multipliers from Block~A to refine adjacent transmission ordering and transmit powers.

1)~\emph{Transmission order:} Given a tightness threshold $\epsilon_{\mathrm{swap}}>0$, consider adjacent transmission endpoints from distinct robots satisfying $\Delta_\ell^\beta\le\epsilon_{\mathrm{swap}}$ and $\mu_\ell^{\mathrm{ord},\star}>10^{-8}$. In decreasing order of $\mu_\ell^{\mathrm{ord},\star}$, we swap disjoint endpoint pairs that preserve $(\prec_g,\prec_d)$ and hence $\mathcal{D}_i^\beta$, yielding a candidate branch $\beta^{\mathrm{cand}}$. An exchange at $\Delta_\ell^\beta=0$ preserves integrated service at the shared boundary; for $\Delta_\ell^\beta>0$, we recompute LMMSE rates and re-optimize the schedule on $\beta^{\mathrm{cand}}$.

2)~\emph{Dual-guided power projection:} Adjusting $p_e$ modifies both task $e$'s SINR and co-channel interference on co-scheduled tasks $j\in\mathcal{A}_\ell\setminus\{e\}$. At $\omega=0$ where $\hat{h}_j^0=h_j$, applying the matrix inversion lemma to $\gamma_j^0=p_j h_j^H C_{-j}^{-1}h_j$ gives the exact sensitivity
\begin{equation}
\frac{\partial\gamma_j^0}{\partial p_e}=\begin{cases}
h_j^H C_{-j}^{-1}h_j>0, & e=j,\\[2pt]
-p_j\big|h_e^H C_{-j}^{-1}h_j\big|^2\le 0, & e\ne j.
\end{cases}
\label{eq:sinr_deriv}
\end{equation}
Combining~\eqref{eq:sinr_deriv} with perturbed scenarios $\omega\in\Omega_{\mathrm{tr}}$, differentiating~\eqref{eq:sinr} via $\partial w_j^\omega/\partial p_e=-(\Sigma_{\mathcal{A}_\ell}^\omega)^{-1}\hat{h}_e^\omega(\hat{h}_e^\omega)^H w_j^\omega$ gives $\partial S_j^{\omega,\beta}/\partial p_e=\sum_{\ell:j,e\in\mathcal{A}_\ell}\Delta_\ell^\beta\frac{\chi W/\ln 2}{1+\gamma_{j\ell}^\omega}\frac{\partial\gamma_{j\ell}^\omega}{\partial p_e}$, where $\gamma_{j\ell}^\omega$ is task $j$'s SINR on interval $\ell$ in scenario $\omega$. Including battery multipliers $\tilde{\nu}_{k_e}^\star$ yields $\mathbf{G}^{(m)}=\{G_e^{(m)}\}_{e\in\mathcal{E}}$:
\begin{equation}
G_e^{(m)}=\zeta\sum_{j,\omega}\mu_{j,\omega}^\star\frac{\partial S_j^{\omega,\beta}}{\partial p_e}-\tilde{\nu}_{k_e}^\star\frac{f_e^{(m+1)}-s_e^{(m+1)}}{\eta_{k_e}^{\mathrm{PA}}}.
\label{eq:power_grad}
\end{equation}
With $\kappa_e^{(m)}=p_{k_e}^{\mathrm{cir}}+p_e^{(m)}/\eta_{k_e}^{\mathrm{PA}}$, let $\Delta\hat{B}_k^{(m)}=\hat{B}_k^{(m)}-\sum_{e\in\pi_k}(\epsilon_k\Gamma_{i_e}(\rho_e^{(m+1)})+\allowbreak\kappa_e^{(m)}(f_e^{(m+1)}-s_e^{(m+1)}))\ge 0$ denote robot $k$'s battery slack, $\mathcal{P}_B^{(m)}=\{\mathbf{p}\in[0_P,\mathbf{p}_{\max}]:\allowbreak\sum_{e\in\pi_k}(f_e^{(m+1)}-s_e^{(m+1)})\allowbreak(p_e-p_e^{(m)})/\eta_k^{\mathrm{PA}}\le\Delta\hat{B}_k^{(m)},\allowbreak\forall k\}$ the battery polytope with $\mathbf{p}_{\max}=\{p_{\max,k_e}\}_{e\in\mathcal{E}}$, $\eta_p>0$ the base step size, and $\epsilon_0>0$ a damping constant. Block~B computes
\begin{equation}
\mathbf{p}^{\mathrm{cand}}=\mathrm{Proj}_{\mathcal{P}_B^{(m)}}\!\left(\mathbf{p}^{(m)}+\eta_p\frac{\mathbf{G}^{(m)}}{\|\mathbf{G}^{(m)}\|_\infty+\epsilon_0}\right),
\label{eq:power_update}
\end{equation}
testing $\{1,0.4,0.1\}\eta_p$ and zero step if $\beta^{\mathrm{cand}}\ne\beta$. Recomputing rates and re-solving Block~A on $\beta^{\mathrm{cand}}$ advances completions; if no trial improves utility in $\mathcal{F}_{\mathrm{tr}}$, Block~B keeps $(\mathbf{p}^{(m)},\beta)$ and Block~A's iterate.

\subsection{Decoupled Sensing--Transmission Waypoint Optimization}
\label{subsec:blockC}
Holding $\boldsymbol{\rho}^{(m+1)}$ and $(s_e^{(m+1)},f_e^{(m+1)})$ fixed, we introduce capture-advance slack $\boldsymbol{\xi}=\{\xi_e\}_{e\in\mathcal{E}}\ge 0_P$ via $g_e(\xi_e)=g_e^{(m+1)}-\xi_e\ge 0$ in $\mathcal{G}_\beta$ preserving $\prec_g$ and $g_e\ge 0$, so $x_e^{\mathrm{tx}}$ may step beyond $v_{k_e}(s_e^{(m+1)}-g_e^{(m+1)})$ when channel gains outweigh age growth. With $\alpha_e^{(m)}=(q_{0,i_e}/d_{0,i_e}^2)/\allowbreak(1+\allowbreak\|x_e^{\mathrm{s},(m)}-\tilde{y}_{i_e}\|^2/d_{0,i_e}^2)^2$, expanding $q_e^{\mathrm{sp}}(x_e^{\mathrm{s}})$ gives the concave quadratic minorant
\begin{multline}
\hat{q}_e^{\mathrm{sp}}(x_e^{\mathrm{s}};x_e^{\mathrm{s},(m)})=q_e^{\mathrm{sp}}(x_e^{\mathrm{s},(m)})\\
-\alpha_e^{(m)}\big(\|x_e^{\mathrm{s}}-\tilde{y}_{i_e}\|^2-\|x_e^{\mathrm{s},(m)}-\tilde{y}_{i_e}\|^2\big),
\label{eq:q_minorant}
\end{multline}
satisfying $q_e^{\mathrm{sp}}(x_e^{\mathrm{s}})e^{-\xi_e/\tau_{i_e}}\ge\hat{q}_e^{\mathrm{sp}}(x_e^{\mathrm{s}};x_e^{\mathrm{s},(m)})-(q_{0,i_e}/\tau_{i_e})\xi_e$ globally with equality at $(x_e^{\mathrm{s},(m)},0)$. Since $\|x_e^{\mathrm{s},(m)}-\tilde{y}_{i_e}\|\ge d_{\min,i_e}+r_y>0$, we linearize the inner sensing boundary via outward unit normal $n_e^{(m)}=(x_e^{\mathrm{s},(m)}-\tilde{y}_{i_e})/\|x_e^{\mathrm{s},(m)}-\tilde{y}_{i_e}\|$. Let $\mathcal{N}_e^\beta$ collect task $e$ and its co-scheduled neighbors with $e\in\mathcal{N}_e^\beta$, $\delta_x>0$ denote a trust radius, $\Delta x_j^{\mathrm{tx}}=x_j^{\mathrm{tx}}-x_j^{\mathrm{tx},(m)}$, and $\mathbf{v}=\{v_e\}_{e\in\mathcal{E}}$ satisfy $v_j\ge\|\Delta x_j^{\mathrm{tx}}\|^2$. On $\|x_e^{\mathrm{tx}}-x_e^{\mathrm{tx},(m)}\|\le\delta_x$, the second-order service surrogate is
\begin{multline}
\underline{S}_e^\omega(\mathbf{x}^{\mathrm{tx}},\mathbf{v})=S_e^\omega(\mathbf{x}^{\mathrm{tx},(m)})\\
+\!\!\sum_{j\in\mathcal{N}_e^\beta}\!\!\Big(\nabla_{x_j^{\mathrm{tx}}} S_e^\omega(\mathbf{x}^{\mathrm{tx},(m)})^\top\Delta x_j^{\mathrm{tx}}-\tfrac{L_{S,e}^\omega}{2}v_j\Big).
\label{eq:s_minorant}
\end{multline}
Here $L_{S,e}^\omega\ge 0$ is $1.5$ times the maximum of $2[\tilde{S}_e^\omega-S_e^\omega]_+/\sum_{j\in\mathcal{N}_e^\beta}\|\Delta x_j^{\mathrm{tx}}\|^2$ over $C_{\mathrm{cur}}=4$ planar perturbations of radius $0.4\delta_x$ around $\tilde{S}_e^\omega$, checking~\eqref{eq:reserve} before acceptance.

Let $L_e^{\mathrm{pre}}(\mathbf{x})=\|x_e^{\mathrm{s}}-x_{e^-}^{\mathrm{tx}}\|$, $L_e^{\mathrm{tr}}(\mathbf{x})=\|x_e^{\mathrm{tx}}-x_e^{\mathrm{s}}\|$, and $\Delta\mathbf{x}=\{x_e^{\mathrm{tx}}-\allowbreak x_e^{\mathrm{s}}\}_{e\in\mathcal{E}}$. With Block-A duals set to zero if unavailable so $\tilde{\nu}_k^\star=\eta_E$, and weights $\eta_x,\eta_{\Delta x},\eta_v>0$, set $\mathcal{E}_{\mathrm{env}}^{(m)}(\mathbf{x},\mathbf{v})=\zeta\sum_{e,\omega}\mu_{e,\omega}^\star(\underline{S}_e^\omega-\allowbreak S_e^{\omega,(m)})-\sum_e(\theta_e^{\mathrm{pre},\star}\allowbreak L_e^{\mathrm{pre}}+\allowbreak\theta_e^{\mathrm{tr},\star}L_e^{\mathrm{tr}})/v_{k_e}-\sum_k\tilde{\nu}_k^\star c_k^{\mathrm{mov}}\sum_{e\in\pi_k}(L_e^{\mathrm{pre}}+\allowbreak L_e^{\mathrm{tr}})$ and $\mathcal{R}_{\mathrm{prox}}^{(m)}=(\eta_x/2)\|\mathbf{x}^{\mathrm{s}}-\mathbf{x}^{\mathrm{s},(m)}\|^2+(\eta_{\Delta x}/2)\|\Delta\mathbf{x}-\Delta\mathbf{x}^{(m)}\|^2+\eta_v\sum_e v_e+(\eta_t/2)\|\boldsymbol{\xi}\|^2$, where $\Delta\mathbf{x}=0_{2P}$ for ``Co-located.'' With Block-B utility $U_{\mathrm{B}}^{(m)}$, $B_k^{\mathrm{mov}}=B_k-E_k^{\mathrm{cpu}}-E_k^{\mathrm{rad}}$, and minorant $\tilde{U}_i(\mathbf{x}^{\mathrm{s}},\boldsymbol{\xi})=\sum_{e\in\mathcal{D}_i^\beta}\tilde{\varpi}_e^{(m+1)}[\hat{q}_e^{\mathrm{sp}}(x_e^{\mathrm{s}};x_e^{\mathrm{s},(m)})-\allowbreak(q_{0,i_e}/\tau_{i_e})\xi_e]$ for $\tilde{\varpi}_e^{(m+1)}=\psi_{i_e}(\rho_e^{(m+1)})\allowbreak\tau_{i_e}(e^{A_e^{(m+1)}}-\allowbreak e^{B_e^{(m+1)}})/H$, Block~C solves~\cite{lobo1998applications}:
\begin{equation}
\begin{aligned}
\max_{\mathbf{x},\boldsymbol{\xi}\ge 0_P,\mathbf{v},\tilde{u}}\; &\tilde{u}+\tfrac{\eta_{\mathrm{sec}}}{N}\!\sum_{i\in\mathcal{I}}\tilde{U}_i(\mathbf{x}^{\mathrm{s}},\boldsymbol{\xi})+\mathcal{E}_{\mathrm{env}}^{(m)}(\mathbf{x},\mathbf{v})-\mathcal{R}_{\mathrm{prox}}^{(m)}\\[-2pt]
\text{s.t.}\;\; &U_{\mathrm{B}}^{(m)}-10^{-7}\le\tilde{u}\le\tilde{U}_i(\mathbf{x}^{\mathrm{s}},\boldsymbol{\xi}),\;\forall i,\;\; \mathbf{g}(\boldsymbol{\xi})\in\mathcal{G}_\beta,\\[-1pt]
&\zeta\underline{S}_e^\omega(\mathbf{x}^{\mathrm{tx}},\mathbf{v})\ge b_{i_e}+\rho_e^{(m+1)}D_{i_e},\;\forall e,\omega,\\[-1pt]
&L_e^{\mathrm{pre}}(\mathbf{x})\le v_{k_e}(g_e(\xi_e)-f_{e^-}^{(m+1)}),\;\forall e,\\[-1pt]
&L_e^{\mathrm{tr}}(\mathbf{x})\le v_{k_e}(s_e^{(m+1)}-g_e(\xi_e)),\;\forall e,\\[-1pt]
&\textstyle\sum_{e\in\pi_k}c_k^{\mathrm{mov}}(L_e^{\mathrm{pre}}(\mathbf{x})+L_e^{\mathrm{tr}}(\mathbf{x}))\le B_k^{\mathrm{mov}},\;\forall k,\\[-1pt]
&\|\Delta x_e^{\mathrm{tx}}\|^2\le v_e,\;x_e^{\mathrm{s}},x_e^{\mathrm{tx}}\in\mathcal{X},\;\|\Delta x_e^{\mathrm{tx}}\|\le\delta_x,\\[-1pt]
&\|x_e^{\mathrm{s}}-x_e^{\mathrm{s},(m)}\|\le\delta_x,\;\|x_e^{\mathrm{s}}-\tilde{y}_{i_e}\|\le d_{\max,i_e}-r_y,\\[-1pt]
&(n_e^{(m)})^\top(x_e^{\mathrm{s}}-\tilde{y}_{i_e})\ge d_{\min,i_e}+r_y,\;\forall e.
\end{aligned}
\label{eq:spatial_socp}
\end{equation}
To clarify when in-transit compression beats serial stop-and-compress relocation under $s_e\ge g_e+T_e^{\mathrm{cpu}}+L_e^{\mathrm{tr}}/v_{k_e}$, Proposition~\ref{prop:relocation_threshold} gives the exact delivery threshold.

\begin{proposition}
\label{prop:relocation_threshold}
Fix packet $e$ with capture time $g_e$, payload $b_{i_e}+\rho_e D_{i_e}$, power $p_e$, and reconstruction time $t_{i_e}^{\mathrm{rec}}$, and compare transmitting at $x_e^{\mathrm{s}}$ versus $y\in\mathcal{X}$ with intra-task transit $T_e^{\mathrm{tr}}=\|y-x_e^{\mathrm{s}}\|/v_{k_e}$. Assume a common constant active set of co-transmitters at fixed positions and powers and positive rates $R_e^\omega(z)$ in all scenarios $\omega\in\Omega_{\mathrm{tr}}\cup\{0\}$. Each of the co-located, pipelined, and serial windows must be feasible, start at readiness, and achieve the earliest delivery satisfying~\eqref{eq:reserve} with minimum duration $\tau_e^{\mathrm{tx}}(z)=(b_{i_e}+\rho_e D_{i_e})/(\zeta\min_{\omega\in\Omega_{\mathrm{tr}}\cup\{0\}}R_e^\omega(z))$. With transmission-duration saving $\Delta\tau_e^{\mathrm{tx}}=\tau_e^{\mathrm{tx}}(x_e^{\mathrm{s}})-\tau_e^{\mathrm{tx}}(y)$, the delivery-time advances relative to co-location are
\begin{equation}
\Delta d_{\mathrm{pipe}}=\Delta\tau_e^{\mathrm{tx}}-[T_e^{\mathrm{tr}}-T_e^{\mathrm{cpu}}]_+,\quad
\Delta d_{\mathrm{serial}}=\Delta\tau_e^{\mathrm{tx}}-T_e^{\mathrm{tr}}.
\label{eq:reloc_threshold}
\end{equation}
Hence relocation strictly advances delivery only under pipelining if and only if $[T_e^{\mathrm{tr}}-T_e^{\mathrm{cpu}}]_+<\Delta\tau_e^{\mathrm{tx}}\le T_e^{\mathrm{tr}}$, and pipelining beats serial relocation by $\min\{T_e^{\mathrm{tr}},T_e^{\mathrm{cpu}}\}$.
\end{proposition}
\begin{IEEEproof}
The worst-rate scenario fixes $\tau_e^{\mathrm{tx}}(z)$ satisfying~\eqref{eq:reserve}. Adding readiness delays $T_e^{\mathrm{cpu}}$, $\max\{T_e^{\mathrm{cpu}},T_e^{\mathrm{tr}}\}$, and $T_e^{\mathrm{cpu}}+T_e^{\mathrm{tr}}$ to $g_e+\tau_e^{\mathrm{tx}}(z)+t_{i_e}^{\mathrm{rec}}$ gives~\eqref{eq:reloc_threshold}; $\Delta d_{\mathrm{pipe}}>0\ge\Delta d_{\mathrm{serial}}$ yields the stated region and $\Delta d_{\mathrm{pipe}}-\Delta d_{\mathrm{serial}}=\min\{T_e^{\mathrm{tr}},T_e^{\mathrm{cpu}}\}$.
\end{IEEEproof}

If advancing $d_e$ to $d'_e<d_e$ alters target $i_e$'s active packet only on $[d'_e,d_e)\subseteq[0,H]$ over prior update $j$,~\eqref{eq:inst_value} gives $\Delta U_{i_e}>0\Leftrightarrow q_e e^{g_e/\tau_{i_e}}>q_j e^{g_j/\tau_{i_e}}$, where the right-hand side is $0$ before first delivery. Along $[x_e^{\mathrm{s}},x_{e^+}^{\mathrm{s}}]$, relocating within $v_{k_e}T_e^{\mathrm{cpu}}$ preserves propulsion distance while cutting RF energy when $\Delta\tau_e^{\mathrm{tx}}>0$.

\subsection{Initial Feasible Schedule Design}
\label{subsec:init}
For $K\le N$, we construct $\widehat{\Pi}$ and seek $z_1=(\widehat{\Pi},\allowbreak\mathbf{x}^{(0)},\allowbreak\boldsymbol{\rho}^{(0)},\allowbreak\mathbf{p}^{(0)},\allowbreak\mathbf{t}^{(0)})\in\mathcal{F}_{\mathrm{tr}}$ in three deterministic steps~\cite{mei2006deployment}:
\begin{enumerate}\setlength{\itemsep}{-0.6pt}
\item \emph{Target assignment:} Targets $\{\tilde{y}_i\}$ sorted by azimuth around $b$ form $K$ contiguous sectors $\mathcal{S}_k$, where the first $N\bmod K$ have size $\lceil N/K\rceil$ and the rest $\lfloor N/K\rfloor$, with depot $x_k^0$ aligned from $b$ toward $\mathcal{S}_k$'s mean estimate.
\item \emph{Tours and revisit counts:} A nearest-neighbor tour on $\mathcal{S}_k$ from $x_k^0$ is refined by 2-opt. Testing $1$--$19$ tours at reference $(\rho,p)$ sets $c_k^{\max}\ge 1$ within reserved $(H,B_k)$ fractions, or $1$ if exceeded, and $c_k=\min\{c_k^{\max},1+\min_j c_j^{\max}\}$ so $\bar{n}_i(\widehat{\Pi})=|\mathcal{E}_i|$ differ by at most one.
\item \emph{Feasible schedule initialization:} For $\tilde{y}_i\ne b$ and $d_{\min,i}+r_y\le d_{\mathrm{init}}\le d_{\max,i}-r_y$, let $e_i^{\mathrm{rad}}=(\tilde{y}_i-b)/\|\tilde{y}_i-b\|$ and set $x_i^{\mathrm{tan}}=\tilde{y}_i+d_{\mathrm{init}}[-e_{i,2}^{\mathrm{rad}},\allowbreak e_{i,1}^{\mathrm{rad}}]^\top\in\mathcal{X}$. In Section~\ref{subsec:ckm_eval}, adaptive CKM and blind schemes replace $x_i^{\mathrm{tan}}$ by maximizing single-link rate at $p_{\max,k}$, breaking ties by smaller $\|x\|$, over $24$ uniform azimuths on $\|x-\tilde{y}_i\|=d_{\min,i}+r_y$ in $\mathcal{X}$, succeeding on $20/20$ and $7/20$ CKM rings at $131$ and $384$~kbits. Over $\mathcal{G}_\rho=\{0.35,\allowbreak 0.5,\allowbreak 0.7,\allowbreak 0.85,\allowbreak 1\}\cap[\rho_{\min,i},1]$ and $\mathcal{G}_p=\{0.02,\allowbreak 0.035,\allowbreak 0.05,\allowbreak 0.1,\allowbreak 0.2,\allowbreak 0.35,\allowbreak 0.5,\allowbreak 0.75,\allowbreak 1\}\cap(0,p_{\max,k}]$ in W, as-soon-as-ready and time-division schedules run, dilating idle gaps with $\mathcal{A}_\ell=\emptyset$ on $[0,H]$ and keeping the best $z_1\in\mathcal{F}_{\mathrm{tr}}$, or repeating after a minimum-travel co-located SOCP on the $1$~mm-tightened BS-facing inner half-space and outer ring of~\eqref{eq:tight_sensing} and chord $x_e^{\mathrm{tx}}$ fallback.
\end{enumerate}

\subsection{Overall Algorithm, Complexity, and Performance Analysis}
\label{subsec:convergence}

\begin{algorithm}[t]
\caption{Alternating Spatial--Resource BCD--SCA}
\label{alg:proposed}
\footnotesize
\begin{algorithmic}[1]
\STATE \textbf{Input:} Feasible $z_1$ from Section~\ref{subsec:init}, scenarios $\Omega_{\mathrm{tr}}\cup\{0\}$, radius $\delta_x$, weights, tolerances $\epsilon_{\mathrm{swap}},\varepsilon_{\mathrm{tol}}>0$, and limit $L_{\max}$.
\STATE Initialize transmission branch $\beta$ from $z_1$, set $m=0$, and compute $U^{(0)}$.
\REPEAT
  \STATE Evaluate rates at $\mathbf{x}^{\mathrm{tx},(m)}$. Solve~\eqref{eq:sub_Qt}, or~\eqref{eq:sub_Qt}--\eqref{eq:sub_Q} for general $\psi_i$; keep a feasible solution if $U\ge U^{(m)}$, else retain the iterate.
  \STATE Set $\beta^{\mathrm{cand}}\leftarrow\beta$; swap disjoint inter-robot pairs with $\Delta_\ell^\beta\le\epsilon_{\mathrm{swap}}$ and $\mu_\ell^{\mathrm{ord},\star}>10^{-8}$ keeping $(\prec_g,\prec_d)$.
  \STATE Test $\mathbf{p}^{\mathrm{cand}}$ in~\eqref{eq:power_update} on $\beta^{\mathrm{cand}}$ via Step~4; accept if in $\mathcal{F}_{\mathrm{tr}}$ with higher utility, else retain Step~4's iterate and $\beta$.
  \STATE Build~\eqref{eq:q_minorant}--\eqref{eq:s_minorant} and solve SOCP~\eqref{eq:spatial_socp} for $(\mathbf{x}^{\mathrm{cand}},\boldsymbol{\xi}^{\mathrm{cand}})$.
  \STATE Search along $(\mathbf{x},\boldsymbol{\xi})$ and $\mathbf{x}^{\mathrm{tx}}$-only directions. Accept the first $\mathcal{F}_{\mathrm{tr}}$ step matching or beating Step~6 under exact rates and~\eqref{eq:reserve} and update $\beta$; else retain Step~6's iterate. Set $m\leftarrow m+1$.
\UNTIL{$|U^{(m)}-U^{(m-1)}|\le\varepsilon_{\mathrm{tol}}$ over $3$ steps at $m\ge 4$, or $m=L_{\max}$.}
\STATE If the final iteration updated $\mathbf{x}$, repeat Steps~4--6 once at those waypoints.
\STATE \textbf{Output:} Optimized feasible solution $(\mathbf{x}^\star,\boldsymbol{\rho}^\star,\mathbf{p}^\star,\mathbf{t}^\star)$.
\end{algorithmic}
\end{algorithm}

\subsubsection{Convergence Analysis}
Algorithm~\ref{alg:proposed} summarizes the overall procedure.
\begin{theorem}
\label{thm:convergence}
Starting from a feasible initial solution $z_1\in\mathcal{F}_{\mathrm{tr}}$, every iterate retained by Algorithm~\ref{alg:proposed} remains feasible for~$(\mathcal{P}[\widehat{\Pi}])$, guarantees nominal service margin at least $(1/\zeta-1)D_e>0$ on~\eqref{eq:service}, and maintains simultaneous true sensing feasibility~\eqref{eq:sensing_area} with probability at least $1-\alpha_y$. Moreover, $\{U^{(m)}\}$ is monotonically non-decreasing and converges to a finite limit $U^\star\le\min_{i\in\mathcal{I}}q_{0,i}$.
\end{theorem}
\begin{IEEEproof}
By induction from $z_1\in\mathcal{F}_{\mathrm{tr}}$, Block~A yields $U\ge U^{(m)}$ on $\mathcal{F}_{\mathrm{tr}}$ via Lemma~\ref{lem:lift_exact}, verified explicitly when $\varepsilon_{\mathrm{fl}}>0$. Block~B projects onto $\mathcal{P}_B^{(m)}$ and accepts only strictly improving $\mathcal{F}_{\mathrm{tr}}$ states, giving $U_{\mathrm{B}}^{(m)}\ge U^{(m)}$. In Block~C, $(n_e^{(m)})^\top(x_e^{\mathrm{s}}-\tilde{y}_{i_e})\ge d_{\min,i_e}+r_y$ implies $\|x_e^{\mathrm{s}}-\tilde{y}_{i_e}\|\ge d_{\min,i_e}+r_y$ by Cauchy--Schwarz, convex travel/battery constraints hold along step sizes in $(0,1]$, and Steps~8 and~10 recompute exact multi-scenario service and accept only $\mathcal{F}_{\mathrm{tr}}$-feasible states with $U\ge U_{\mathrm{B}}^{(m)}$ regardless of whether surrogate~\eqref{eq:s_minorant} is globally conservative, proving $U^{(m+1)}\ge U^{(m)}$. On $\omega=0$,~\eqref{eq:reserve} yields $S_e^0-D_e\ge(1/\zeta-1)D_e>0$, while~\eqref{eq:tight_sensing} ensures~\eqref{eq:sensing_area} with probability at least $1-\alpha_y$, and $U_i\le q_{0,i}$ yields convergence.
\end{IEEEproof}

\begin{remark}
\label{rem:strict}
Theorem~\ref{thm:convergence} assumes exact checks; tolerances $5\times 10^{-5}$ on sensing, domain, causality, and relative service $\delta_{\mathrm{svc}}$, $10^{-6}$ on ratio/power bounds, and $10^{-3}$~J on battery, with nominal and sweep max residuals $9.1\times 10^{-11}$~m, $5.0\times 10^{-8}$~s, and $0$~J, retain $(1-\delta_{\mathrm{svc}})/\zeta-1\approx 25\%$ reserve at $\zeta=0.80$.
\end{remark}

\subsubsection{Complexity Analysis}
Let $S=|\Omega_{\mathrm{tr}}|+1$ be the scenario count, $C_{\mathrm{ch}}=M^3+KM^2+K^2M+K^3$ the per-interval matrix cost, $I_{\mathrm{2opt}}$ the 2-opt pass cap, $B_{\mathrm{bis}}$ the bisection steps in~\eqref{eq:power_update}, and $\varepsilon_{\mathrm{ipm}}>0$ the interior-point accuracy. Building $\widehat{\Pi}$ and $z_1$ takes $\mathcal{O}(N\log N+\allowbreak I_{\mathrm{2opt}}N^3/K^2+\allowbreak|\mathcal{G}_\rho||\mathcal{G}_p|[P\log P+\allowbreak SP(MR_f+C_{\mathrm{ch}})])$ operations, plus $\mathcal{O}(P^{3.5}\allowbreak\log(1/\varepsilon_{\mathrm{ipm}}))$ if the fallback SOCP runs. Each outer iteration of Algorithm~\ref{alg:proposed} has at most $2P-1$ intervals and $K$ active transmitters. Specifically, LMMSE rates and~\eqref{eq:power_grad} cost $\mathcal{O}(SP C_{\mathrm{ch}})$, projection~\eqref{eq:power_update} costs $\mathcal{O}(P B_{\mathrm{bis}})$, $L_{S,e}^\omega$ estimation costs $\mathcal{O}(S C_{\mathrm{cur}}[P(MR_f+C_{\mathrm{ch}})+\allowbreak P^2])$, and SOCP~\cite{lobo1998applications} or fixed-dimension conic subproblems with $\mathcal{O}(P)$ variables and $\mathcal{O}((S+A)P)$ constraints cost $\mathcal{O}((S+A)^{1.5}P^{3.5}\allowbreak\log(1/\varepsilon_{\mathrm{ipm}}))$, giving $\mathcal{O}(L_{\max}P^{3.5}\allowbreak\log(1/\varepsilon_{\mathrm{ipm}}))$ overall for fixed $(K,M,R_f,S,A)$.

\subsubsection{Analytical Feasibility and Utility Bounds}
Conditional on $\widehat{\Pi}$ with $\bar{n}_i=\bar{n}_i(\widehat{\Pi})=|\mathcal{E}_i|\ge 1$, define $\bar{q}_i=q_{0,i}/\allowbreak(1+(d_{\min,i}+r_y)^2/d_{0,i}^2)$, written $\bar{q}_i^{\mathrm{phys}}$ at $r_y=0$, $\ell_{0,i}=t_{0,i}+t_i^{\mathrm{rec}}$, $(F_{\max},v_{\max},p_{\max})=\max_{k\in\mathcal{K}}(F_k,\allowbreak v_k,\allowbreak p_{\max,k})$, $R_i^{\mathrm{in}}=[\|\tilde{y}_i-b\|-\allowbreak(d_{\max,i}-r_y)]_+$, and $r_i(\Delta r)=([R_i^{\mathrm{in}}-\Delta r]_+^2+\allowbreak h_{\mathrm{BS}}^2)^{1/2}$. Let $R_i^{\mathrm{su}}(\Delta r)=\chi W\allowbreak\log_2(1+\allowbreak p_{\max}\bar{\beta}_i^{\mathrm{ub}}(\Delta r)/(N_0 W F_n))$ with $R_i^{\mathrm{su,col}}\triangleq R_i^{\mathrm{su}}(0)$, where $\bar{\beta}_i^{\mathrm{ub}}(\Delta r)$ equals $\beta_0(r_i(\Delta r)/r_0)^{-\nu}$ under homogeneous path loss and uses certified cell-wise distance and pre-$\tanh$ Taylor--Hessian shadowing bounds over outer sensing disks under a CKM; numerically, both evaluate $\ell_i^\star$ conservatively across $128$ uniform $\Delta r$ bins by pairing upper-bin gain with lower-bin transit delay. At fixed waypoints $\mathbf{x}$ with single-link gain $G_e(\mathbf{x}^{\mathrm{tx}})=\|h_e(x_e^{\mathrm{tx}})\|^2>0$ and single-user rate $R_e^{\mathrm{su}}(\mathbf{x}^{\mathrm{tx}})=\chi W\allowbreak\log_2(1+\allowbreak p_{\max,k_e}G_e(\mathbf{x}^{\mathrm{tx}}))$, define minimum energy per bit $e_e^{\mathrm{bit}}(\mathbf{x}^{\mathrm{tx}})=\inf_{p\in(0,p_{\max,k_e}]}(p_{k_e}^{\mathrm{cir}}+\allowbreak p/\eta_{k_e}^{\mathrm{PA}})/\allowbreak[\zeta\chi W\log_2(1+\allowbreak p G_e(\mathbf{x}^{\mathrm{tx}}))]$, breakpoint set $\mathcal{B}_i\subset[\rho_{\min,i},1]$ collecting the endpoints and breakpoints of $\Gamma_i$, minimum packet energy $\mathcal{E}_e^{\mathrm{pkt},\star}(\mathbf{x}^{\mathrm{tx}})=\min_{\rho\in\mathcal{B}_{i_e}}\{E_{0,i_e}+\allowbreak\epsilon_{k_e}\Gamma_{i_e}(\rho)+\allowbreak e_e^{\mathrm{bit}}(\mathbf{x}^{\mathrm{tx}})(b_{i_e}+\rho D_{i_e})\}$, and $A_{\mathrm{tot}}^\star(\mathbf{x})=\sum_{k:\pi_k\ne\emptyset}\eta_k^{\mathrm{PA}}[B_k-\allowbreak c_k^{\mathrm{mov}}L_k(\mathbf{x})-\allowbreak\sum_{e\in\pi_k}E_{0,i_e}]_+\allowbreak\max_{e\in\pi_k}G_e(\mathbf{x}^{\mathrm{tx}})$. Over $\Delta r\in[0,R_i^{\mathrm{in}}]$, or $\Delta r\in[0,v_{\max}H]$ under a CKM, define the decoupled latency floor
\begin{equation}
\ell_i^{\mathrm{svc}}(\rho,\Delta r)=t_i^{\mathrm{rec}}+\max\!\Big\{t_{0,i}+\tfrac{\Gamma_i(\rho)}{F_{\max}},\tfrac{\Delta r}{v_{\max}}\Big\}+\tfrac{b_i+\rho D_i}{\zeta R_i^{\mathrm{su}}(\Delta r)}.
\label{eq:ell_svc}
\end{equation}
Minimizing~\eqref{eq:ell_svc} gives $\ell_i^\star=\allowbreak\min_{\rho,\Delta r}\ell_i^{\mathrm{svc}}(\rho,\Delta r)>\ell_{0,i}$ and $\Upsilon_i^\star=\allowbreak\max_{\rho,\Delta r}\psi_i(\rho)\allowbreak e^{-\ell_i^{\mathrm{svc}}(\rho,\Delta r)/\tau_i}$. For integer $n\ge 1$, set $F_i(n;\ell_i^\star)=\allowbreak\bar{q}_i\Upsilon_i^\star(n\tau_i/H)\allowbreak[1-e^{-(H-\ell_i^\star)/(n\tau_i)}]$ and $F_i^{\mathrm{base}}(n;\ell_{0,i})=\allowbreak\bar{q}_i e^{-\ell_{0,i}/\tau_i}(n\tau_i/H)\allowbreak[1-e^{-(H-\ell_{0,i})/(n\tau_i)}]$, with $F_i(0;\ell_i^\star)=0$ and $F_i^{\mathrm{base}}(0;\ell_{0,i})=0$. At fixed $\mathbf{x}$, let $\underline{\tau}_e(\rho;\mathbf{x})=\allowbreak\max\{t_{0,i_e}+\allowbreak\Gamma_{i_e}(\rho)/F_{k_e},\allowbreak L_e^{\mathrm{tr}}(\mathbf{x})/v_{k_e}\}+\allowbreak(b_{i_e}+\rho D_{i_e})/\allowbreak(\zeta R_e^{\mathrm{su}}(\mathbf{x}^{\mathrm{tx}}))$, $\underline{\ell}_e=t_{i_e}^{\mathrm{rec}}+\underline{\tau}_e$, tour recursion $\underline{f}_e(\mathbf{x})=\allowbreak\underline{f}_{e^-}(\mathbf{x})+\allowbreak L_e^{\mathrm{pre}}(\mathbf{x})/v_{k_e}+\allowbreak\min_\rho\underline{\tau}_e(\rho;\mathbf{x})$ with $\underline{f}_{e^-}=0$ at the depot, $\underline{d}_i(\mathbf{x})=\allowbreak\min_{e\in\mathcal{E}_i}\{\underline{f}_e(\mathbf{x})+\allowbreak t_i^{\mathrm{rec}}\}$, $\bar{Q}_i(\mathbf{x})=\allowbreak\max_{e\in\mathcal{E}_i}q_e^{\mathrm{sp}}(x_e^{\mathrm{s}})$, and $\Upsilon_i(\mathbf{x})=\allowbreak\max_{e\in\mathcal{E}_i,\rho}\psi_i(\rho)\allowbreak e^{-\underline{\ell}_e(\rho;\mathbf{x})/\tau_i}$, giving $U_{\mathbf{x}}^+\triangleq\allowbreak\min_{i\in\mathcal{I}}\bar{Q}_i(\mathbf{x})\allowbreak\Upsilon_i(\mathbf{x})(\bar{n}_i\tau_i/H)\allowbreak[1-e^{-[H-\underline{d}_i(\mathbf{x})]_+/(\bar{n}_i\tau_i)}]$.

\begin{proposition}
\label{prop:upper_bound}
By Appendix~\ref{app:proof_prop3}, every feasible solution of~$(\mathcal{P}[\widehat{\Pi}])$ obeys:

1)~\emph{Battery, single-stream, and spatial-DoF limits:} At fixed $\mathbf{x}$ with effective multiplexing rank $M_{\mathrm{eff}}=\min\{K,M\}$ evaluated at $\omega=0$,
\begin{gather}
B_k\ge B_{k,\mathrm{crit}}(\mathbf{x})\triangleq c_k^{\mathrm{mov}}L_k(\mathbf{x})+\sum_{e\in\pi_k}\mathcal{E}_e^{\mathrm{pkt},\star}(\mathbf{x}^{\mathrm{tx}}),\;\forall k,\label{eq:bat_crit}\\
\sum_{e\in\mathcal{E}}\frac{b_{i_e}+\rho_e D_{i_e}}{\zeta}\le\chi W M_{\mathrm{eff}} H\log_2\!\left(1+\frac{A_{\mathrm{tot}}^\star(\mathbf{x})}{M_{\mathrm{eff}} H}\right).\label{eq:dof_cap}
\end{gather}
Under co-located single-stream service with $|\mathcal{A}(t)|\le 1$ and uniform $(D,b,\bar{\rho})$ at common ratio $\bar{\rho}$, feasibility requires $D\le D_{\mathrm{TD}}^{\mathrm{col}}(\bar{\rho})\triangleq[H/\allowbreak(\sum_i\bar{n}_i/(\zeta R_i^{\mathrm{su,col}}))-\allowbreak b]/\bar{\rho}$.

2)~\emph{Plan- and waypoint-conditioned utility bounds:} With $n_i=|\mathcal{D}_i|\le\bar{n}_i$,
\begin{align}
U_i&\le F_i(n_i;\ell_i^\star)\le F_i^{\mathrm{base}}(n_i;\ell_{0,i}),\label{eq:prop2_bound}\\
U&\le U_{\mathbf{x}}^+\le U_{\mathrm{UB}}^{\mathrm{rate}}(\widehat{\Pi})\triangleq\min_{i\in\mathcal{I}}F_i(\bar{n}_i;\ell_i^\star)\le U_{\mathrm{UB}}(\widehat{\Pi}),\label{eq:system_ub}
\end{align}
where $U_{\mathrm{UB}}(\widehat{\Pi})=\min_{i\in\mathcal{I}} F_i^{\mathrm{base}}(\bar{n}_i;\ell_{0,i})$, $\Delta r=0$ gives $U_{\mathrm{UB}}^{\mathrm{rate,col}}(\widehat{\Pi})$, $\bar{q}_i\leftarrow\bar{q}_i^{\mathrm{phys}}$ gives the universal out-of-sample bound $U_{\mathrm{UB}}^{\mathrm{phys}}(\widehat{\Pi})$, and $(\bar{q}_i,\zeta,\bar{\beta}_i^{\mathrm{ub}})\leftarrow(\bar{q}_i^{\mathrm{phys}},1,\sup_{x\in\mathcal{X}}\bar{\beta}(x))$ gives $U_{\mathrm{UB}}^{\mathrm{phys,rate}}(\widehat{\Pi})$ for deterministic channels with $\beta_e^{\mathrm{phy}}\le\sup_x\bar{\beta}(x)$ and $K_{\mathrm{R}}\to\infty$, holding for any $\epsilon_{\mathrm{CSI}}\ge 0$ via Cauchy--Schwarz.
\end{proposition}

\section{Simulation Results}
\label{sec:results}

\subsection{Simulation Setup}

\begin{table*}[t]
\centering
\caption{Main System Parameters}
\label{tab:params}
\footnotesize
\setlength{\tabcolsep}{2.8pt}
\renewcommand{\arraystretch}{0.84}
\begin{tabular}{@{}ll@{\hspace{5pt}}ll@{\hspace{5pt}}ll@{}}
\toprule
\textbf{Parameter} & \textbf{Value} & \textbf{Parameter} & \textbf{Value} & \textbf{Parameter} & \textbf{Value} \\
\midrule
Mobile robots $K$, targets $N$ & $8,\;30$ & BS antennas $M$, height $h_{\mathrm{BS}}$ & $8,\;10\,\text{m}$ & Mission duration $H$, speed $v_k$ & $600\,\text{s},\;2\,\text{m/s}$ \\
Sensing ring $[d_{\min,i},d_{\max,i}]$ & $[5,10]\,\text{m}$ & Radial margin $r_y$, $\sigma_y,\alpha_y$ & $0.20\,\text{m},\;0.05\,\text{m},\;0.01$ & Reference $q_{0,i}$, scale $d_{0,i}$ & $1.0,\;15\,\text{m}$ \\
Freshness constant $\tau_i$ & $30\,\text{s}$ & Feature size $D_i$, header $b_i$ & $131\,\text{kbits},\;512\,\text{bits}$ & Compression floor $\rho_{\min,i}$ & $0.35$ \\
CPU clock $F_k$, dynamic $\epsilon_k$ & $1\,\text{GHz},\;2\,\text{nJ/cyc}$ & Workload scale $c_i$ & $6.0\times 10^8\,\text{cycles}$ & Base CPU $t_{0,i},E_{0,i}$, BS $t_i^{\mathrm{rec}}$ & $0.5\,\text{s},\;2\,\text{J},\;0.05\,\text{s}$ \\
Bandwidth $W$, factor $\chi$ & $4\,\text{kHz},\;0.80$ & Noise $N_0,F_n$ in~\eqref{eq:channel} & $-174\,\text{dBm/Hz},\;7\,\text{dB}$ & Path gain $\beta_0$, exponent $\nu$ & $6.3\times 10^{-5},\;3.2$ \\
Service reserve $\zeta$, $\epsilon_{\mathrm{CSI}}$ & $0.80,\;10^{-4}$ & Battery $B_k$, propulsion $c_k^{\mathrm{mov}}$ & $1000\,\text{J},\;1.5\,\text{J/m}$ & Power $p_{\max,k},p_k^{\mathrm{cir}}$, PA $\eta_k^{\mathrm{PA}}$ & $1\,\text{W},\;1\,\text{W},\;0.35$ \\
\bottomrule
\end{tabular}
\vspace{-2pt}
\end{table*}

\begin{figure*}[t]
\centering
\includegraphics[width=.91\textwidth]{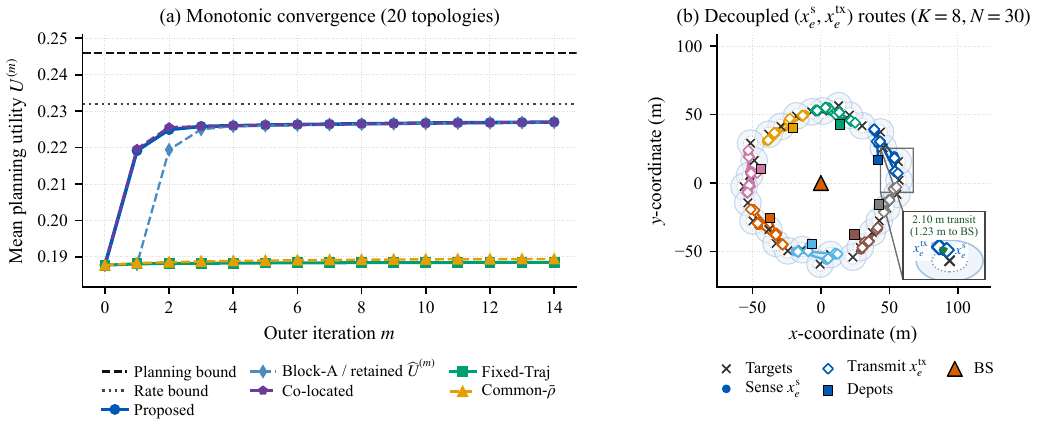}
\caption{Mean planning utility over 20 topologies and optimized waypoints on one topology; inset shows mean target-1 visits shifting over $0.2$~m.}
\label{fig:conv_traj}
\end{figure*}

We consider $K=8$ robots and an $M=8$-antenna BS with $\delta=0.5$ at $b=(0,0)$ in $\mathcal{X}=\{x\in\mathbb{R}^2:\|x\|\le 100\,\text{m}\}$~\cite{mei2006deployment,zhao2025energy,bjornson2015optimal}. Across $20$ random topologies, depots sit at $45$~m and $N=30$ targets lie around a $55$~m ring under random rotation and $\pm 0.055$~rad, $\pm 4.5$~m perturbations, with $d_{\mathrm{init}}=8.8$~m and fixed $\widehat{\Pi}$ in Table~\ref{tab:params}. Following~\cite{zhao2025energy,zhu2025scene},~\eqref{eq:workload} uses $(\varrho_1,\varrho_2,\rho_{\min,i})=(0.70,0.50,0.35)$, $(\varsigma_1,\varsigma_2,\varsigma_3)=(1,3,8)$, $c_i\propto D_i$, and $\psi_i\equiv 1$ unless noted. Planning uses $\omega=0$ and $|\Omega_{\mathrm{tr}}|=32$ realizations with $R_f=32$ and $L_{\mathrm{corr}}=15$~m; evaluation uses $512$ shared realizations fixing $(\tilde{y}_i,\widehat{\Pi})$ and resampling $y_i^{(w)}\sim\mathcal{N}(\tilde{y}_i,\sigma_y^2 I_2)$, with $K_{\mathrm{R}}\to\infty$ by default, $K_{\mathrm{R}}\in\{0,0\,\text{dB},3\,\text{dB},6\,\text{dB},10\,\text{dB}\}$ in~\eqref{eq:rician_phy}, and ray-blocked $K_{\mathrm{R}}$ in Section~\ref{subsec:ckm_eval}. Delivered utility averages over realizations and topologies, scoring infeasible topologies zero, while energy and $\bar{\rho}_{\mathrm{pkt}}$ average over feasible topologies. We compare ``Proposed'' with four benchmark groups:
\begin{itemize}\setlength{\itemsep}{-0.6pt}
\item \textbf{Waypoint decoupling:} ``Co-located'' sets $x_e^{\mathrm{tx}}=x_e^{\mathrm{s}}$; ``Serial-Decoupled'' enforces $s_e\ge g_e+T_e^{\mathrm{cpu}}+L_e^{\mathrm{tr}}/v_{k_e}$.
\item \textbf{Trajectory and compression:} At tangential waypoints, ``Fixed-Traj'' optimizes $(\boldsymbol{\rho},\mathbf{p},\mathbf{t})$, ``Common-$\bar{\rho}$'' enforces uniform $\bar{\rho}$, and ``Full-Feature'' fixes $\rho_e=1$.
\item \textbf{Power and scheduling:} At tangential waypoints, equal power allocation (EPA) fixes uniform power and optimizes $(\bar{\rho},\mathbf{t})$, non-convex local search (NLS) searches discrete $(\bar{\rho},\mathbf{p},\mathbf{t})$, signal-to-noise ratio (SNR) rule ``SNR-only $\rho$'' maps single-link gains to $\boldsymbol{\rho}$ before optimizing $(\mathbf{p},\mathbf{t})$, and ``Oldest-First'' serves the stalest target.
\item \textbf{Boundary rings:} ``Outer-Boundary'' and ``Inner-Boundary'' fix co-located sites on the $2$~mm-inset outer and BS-facing inner sensing rings and optimize $(\boldsymbol{\rho},\mathbf{p},\mathbf{t})$.
\end{itemize}
\begin{figure*}[t]
\centering
\includegraphics[width=.91\textwidth]{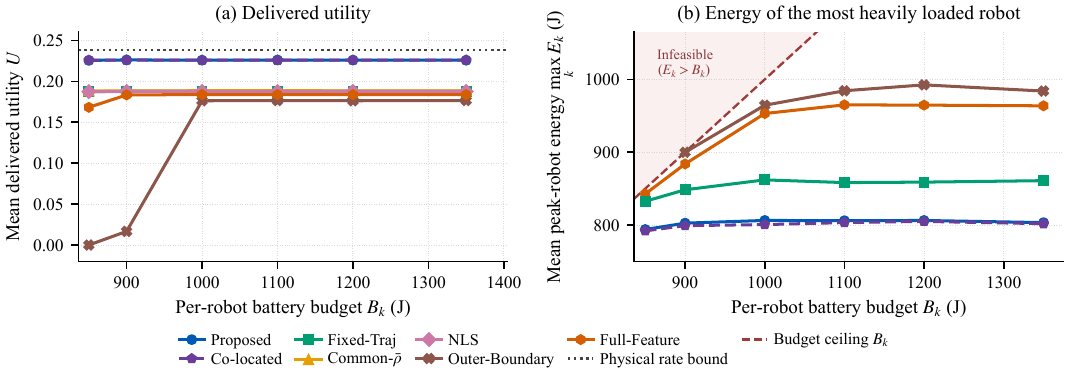}
\caption{Effect of the per-robot battery budget on mean delivered utility and on the energy consumed by the most heavily loaded robot.}
\label{fig:battery_sweep}
\end{figure*}

\begin{figure*}[t]
\centering
\includegraphics[width=.91\textwidth]{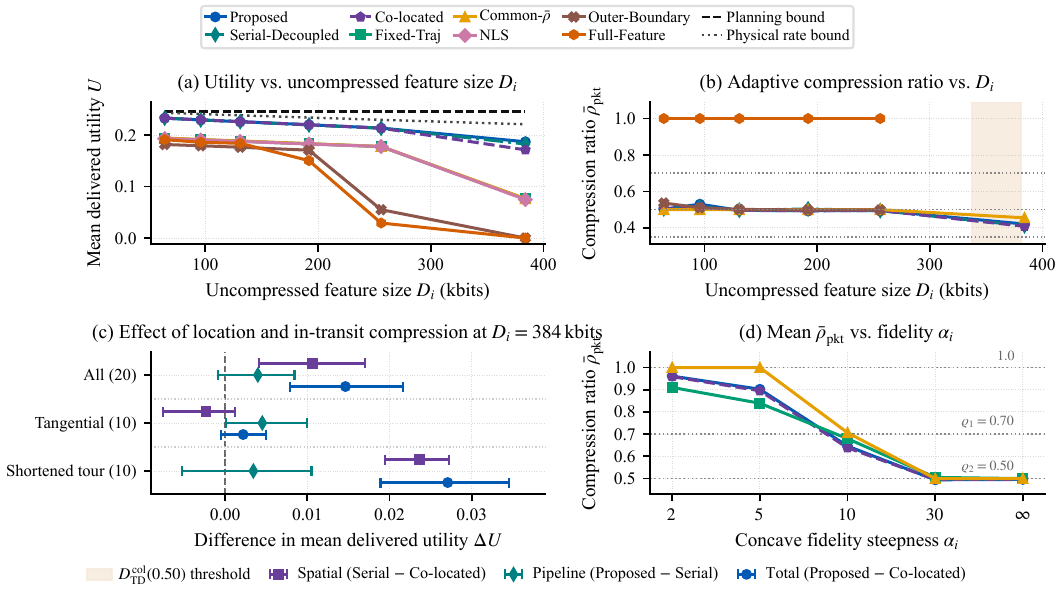}
\caption{Effects of payload and fidelity on utility and compression; ``Planning bound'' bounds planning utility and bars in~(c) show paired $95\%$ bootstrap CIs.}
\label{fig:param_sweeps}
\end{figure*}

\subsection{Convergence and Energy Consumption}

In Fig.~\ref{fig:conv_traj}(a), ``Proposed'' raises mean planning utility from $0.1877$ to $0.2249$ in two steps and $0.2270$ in $14$, reaching $97.9\%$ of $\widehat{\Pi}$'s $0.2320$ $128$-bin rate bound, versus $0.1885$ and $0.1894$ for ``Fixed-Traj'' and ``Common-$\bar{\rho}$,'' while $U_{\mathbf{x}^{(0)}}^+$ averages $0.1911$. Across $20$ topologies, one shown in Fig.~\ref{fig:conv_traj}(b), waypoint control cuts mean sensing range from $8.80$ to $6.08$~m with a $1.16$~m transmit shift and travel by $258.6$~m over ``Fixed-Traj.''

\begin{table}[t]
\centering
\caption{Mean delivered utility, compression ratio, and swarm energy in J at $D_i=131$~kbits and $B_k=1000$~J}
\label{tab:comparison}
\footnotesize
\setlength{\tabcolsep}{3pt}
\renewcommand{\arraystretch}{1.05}

\begin{tabular*}{\columnwidth}{@{\extracolsep{\fill}}lrrr@{}}
\toprule
\textbf{Scheme} & \textbf{Utility} & $\boldsymbol{\bar{\rho}_{\mathrm{pkt}}}$ & \textbf{Energy} \\
\midrule
Proposed & $\mathbf{0.2259}$ & $0.496$ & $5625.4$ \\
Co-located & $\mathbf{0.2259}$ & $0.497$ & $5594.8$ \\
Oldest-First & $0.1892$ & $0.510$ & $5754.9$ \\
Common-$\bar{\rho}$ & $0.1886$ & $0.500$ & $6047.4$ \\
Fixed-Traj & $0.1879$ & $0.500$ & $6110.5$ \\
EPA & $0.1879$ & $0.505$ & $6152.2$ \\
NLS & $0.1877$ & $0.540$ & $6175.8$ \\
SNR-only $\rho$ & $0.1853$ & $0.669$ & $6356.7$ \\
Full-Feature & $0.1839$ & $1.000$ & $6750.9$ \\
Outer-Boundary & $0.1763$ & $0.501$ & $6794.1$ \\
\bottomrule
\end{tabular*}

\end{table}

Table~\ref{tab:comparison} summarizes nominal results. Under equal budgets, ``Proposed'' outperforms ``Fixed-Traj'' by $20.2\%$, saving $485.1$~J with $387.8$~J in propulsion, outperforms ``Outer-Boundary'' by $28.2\%$ with $17.2\%$ lower energy, and saves $16.7\%$ over ``Full-Feature.'' ``Co-located'' also reaches $0.2259$, while ``Inner-Boundary'' attains $0.2257$ at $5426.3$~J, $204.6$~J below re-solved ``Proposed'' at $5630.9$~J. In Fig.~\ref{fig:battery_sweep}, ``Proposed'' and ``Co-located'' stay feasible on $20/20$ rings at $B_k=850$~J with peak loads $793.8$ and $791.9$~J, whereas~\eqref{eq:bat_crit} for ``Outer-Boundary'' spans $897.0$--$963.1$~J, excluding all $20$ rings at $850$~J and $18$ at $900$~J.

\subsection{Effects of Payload, Geometry, and Semantic Fidelity}

Figs.~\ref{fig:param_sweeps}(a) and~\ref{fig:param_sweeps}(b) test $D_i\in[64,384]$~kbits. For $D_i\le 256$~kbits, adaptive-waypoint utilities differ by at most $5\times 10^{-4}$ as $c_i\propto D_i$ keeps $\bar{\rho}_{\mathrm{pkt}}$ near $\varrho_2=0.50$. At $384$~kbits, $D_i$ exceeds Proposition~\ref{prop:upper_bound}'s single-stream fixed-$\bar{\rho}=0.50$ co-located limit $D_{\mathrm{TD}}^{\mathrm{col}}(0.50)$ of $337.8$--$381.1$~kbits on all $20$ rings; initialization finds no ``Full-Feature'' schedule on $0/20$ rings, ``Co-located'' solves $20/20$ rings at $\bar{\rho}_{\mathrm{pkt}}=0.409$ by compressing below $0.50$, and ``Proposed'' beats ``Co-located'' by $9.2\%$ at $0.1870$ versus $0.1712$.

Fig.~\ref{fig:param_sweeps}(c) and Table~\ref{tab:r1_attribution} separate relocation from transit compression at $384$~kbits from a common start. Over ``Co-located'' at $0.1719$, ``Serial-Decoupled'' adds $0.0107$ and transit overlap adds $0.0040$ whose $95\%$ CI includes zero. Together they yield an $8.5\%$ gain at $7.1\%$ higher energy from $6161.4$ to $6598.5$~J, or $0.42\%$ when also selecting ``Inner-Boundary.'' Gains concentrate on the $10$ shortened-tour rings; under nested warm starts, ``Proposed'' reaches $0.1922$ at $6502.6$~J versus $0.1930$ for ``Inner-Boundary'' at $6959.0$~J, saving $456.4$~J.

\begin{table}[t]
\centering
\caption{Mean delivered utility at $384$~kbits from a common start unless noted}
\label{tab:r1_attribution}
\footnotesize
\setlength{\tabcolsep}{2pt}
\renewcommand{\arraystretch}{1.05}

\begin{tabular*}{\columnwidth}{@{\extracolsep{\fill}}lrrr@{}}
\toprule
\textbf{Setting} & \textbf{Proposed} & \textbf{Serial-Decoupled} & \textbf{Co-located} \\
\midrule
Ring, LoS & $\mathbf{0.1866}$ & $0.1826$ & $0.1719$ \\
\quad Tangential, $n=10$ & $\mathbf{0.1823}$ & $0.1778$ & $0.1801$ \\
\quad Shortened init., $n=10$ & $\mathbf{0.1909}$ & $0.1874$ & $0.1638$ \\
Ring, LoS, nested starts & $\mathbf{0.1922}$ & $0.1873$ & $0.1719$ \\
Ring, Rician, $K_{\mathrm{R}}=6$~dB & $\mathbf{0.1866}$ & $0.1826$ & $0.1719$ \\
Ring, Rician, $K_{\mathrm{R}}=0$~dB & $\mathbf{0.1865}$ & $0.1825$ & $0.1719$ \\
Ring, Rayleigh, $K_{\mathrm{R}}=0$ & $\mathbf{0.1858}$ & $0.1821$ & $0.1714$ \\
Clustered, $K_{\mathrm{R}}\in\{0,6\,\text{dB},\infty\}$ & $\mathbf{0.0768}$ & $0.0767$ & $0.0767$ \\
\bottomrule
\end{tabular*}

\end{table}

Proposition~\ref{prop:relocation_threshold} explains why transit pipelining yields a modest gain in homogeneous LoS channels. Relocating $x_e^{\mathrm{tx}}$ advances delivery only when saved wireless duration exceeds residual travel after compression; at $384$~kbits, $213$ of $235$ constant-interferer windows within $5\times 10^{-5}$ peer-service tolerance advance solely via transit compression, with median positive advance $0.915$~ms versus $\tau_i=30$~s. On clustered topologies with $12,10,8$ targets, ``Proposed'' improves utility by $20.1\%$ over ``Fixed-Traj'' and $0.11\%$ over ``Co-located'' in Table~\ref{tab:r1_attribution}. Under spatially correlated multipath~\eqref{eq:rician_phy} with $R_{\mathrm{rx},e}=I_M$, $\zeta=0.80$, and $L_{\mathrm{corr}}=15$~m on $384$~kbit rings, ``Proposed'' satisfies wireless service and~\eqref{eq:sensing_area} jointly in $10{,}238$, $10{,}232$, $10{,}193$, and $9{,}940$ of $10{,}240$ realizations for $K_{\mathrm{R}}\in\{6\,\text{dB},3\,\text{dB},0\,\text{dB},0\}$, delivering $99.98\%$ of $1{,}916{,}928$ packets at $K_{\mathrm{R}}=0$. At $K_{\mathrm{R}}=0$, it retains $0.1858$ at $384$~kbits, $+8.4\%$ over ``Co-located,'' $0.2259$ over $10{,}230/10{,}240$ tests at $131$~kbits in Table~\ref{tab:comparison}, and $0.0768$ over $10{,}238/10{,}240$ tests on clusters.

As shown in Fig.~\ref{fig:param_sweeps}(d), ``Proposed'' increases $\bar{\rho}_{\mathrm{pkt}}$ from $0.496$ under $\psi_i\equiv 1$ and $0.494$ at $\alpha_i=30$ to $0.961$ at $\alpha_i=2$. This is because steeper fidelity loss raises the marginal utility penalty of pruning in Proposition~\ref{prop:kkt}, shifting $\rho_e^\star$ above the workload breakpoints. Moreover, varying $\tau_i\in[10,90]$~s raises utility from $0.0680$ to $0.4867$, while $M\in[4,16]$ keeps utility at $0.2259$--$0.2260$ under the $1/\sqrt{M}$ array scaling in~\eqref{eq:channel}.

\subsection{Site-Specific Obstacle Shadowing and CKM Evaluation}
\label{subsec:ckm_eval}

\begin{table}[t]
\centering
\caption{Mean delivered utility and end-to-end joint completion rate OK in $\%$ under 2D obstacle shadowing over 20 topologies}
\label{tab:ckm_shadow}
\footnotesize
\setlength{\tabcolsep}{2.4pt}
\renewcommand{\arraystretch}{1.0}
\begin{tabular*}{\columnwidth}{@{\extracolsep{\fill}}lrrrr@{}}
\toprule
& \multicolumn{2}{c}{$\boldsymbol{D_i=131}$~\textbf{kbits}} & \multicolumn{2}{c}{$\boldsymbol{D_i=384}$~\textbf{kbits}} \\
\textbf{Scheme} & \textbf{Utility} & \textbf{OK} & \textbf{Utility} & \textbf{OK} \\
\midrule
CKM-Proposed & $0.2224$ & $93.8$ & $\mathbf{0.1860}$ & $91.5$ \\
CKM-Serial & $0.2212$ & $96.3$ & $0.1778$ & $85.0$ \\
CKM-Co-located & $0.2213$ & $95.7$ & $0.1654$ & $91.6$ \\
CKM-Inner & $0.2200$ & $91.7$ & $0.1711$ & $84.9$ \\
CKM-Fixed & $0.1816$ & $91.1$ & $0.0202$ & $8.9$ \\
Blind-Margin & $\mathbf{0.2233}$ & $99.7$ & $0.1493$ & $50.5$ \\
Blind-Uncal & $0.0638$ & $22.1$ & $0.0349$ & $13.8$ \\
\bottomrule
\end{tabular*}
\end{table}

To evaluate site-specific blockage, $N_{\mathrm{obs}}=9$ cylindrical obstacles lie on the $[24,32]$~m ring with centers $z_o^{\mathrm{obs}}\in\mathbb{R}^2$ under $\pm 0.10$~rad jitter and radii $\varrho_o^{\mathrm{obs}}\in[3.8,4.8]$~m satisfying $\|z_o^{\mathrm{obs}}\|+\varrho_o^{\mathrm{obs}}<40$~m, clear of depots and outer sensing disks, for $o\in\{1,\ldots,N_{\mathrm{obs}}\}$. With $\sigma_{\mathrm{lg}}(z)=(1+e^{-z})^{-1}$, projection $t_o(x)=(x^\top z_o^{\mathrm{obs}})/\allowbreak(\|x\|^2+\varepsilon_0^2)$ at $\varepsilon_0=0.25$~m, gate $g_o(x)=\sigma_{\mathrm{lg}}((t_o(x)-0.06)/0.03)\allowbreak\sigma_{\mathrm{lg}}((0.94-t_o(x))/0.03)$, clearance $d_{\perp,o}(x)=(\|z_o^{\mathrm{obs}}-t_o(x)x\|^2+\allowbreak\varepsilon_\perp^2)^{1/2}$ at $\varepsilon_\perp=0.05$~m, margin $r_{\mathrm{map}}=1.20$~m, signed clearance $\Delta d_o(x)=d_{\perp,o}(x)-\varrho_o^{\mathrm{obs}}-r_{\mathrm{map}}$, and smoothing width $\ell_{\mathrm{sm}}=0.45$~m govern each obstacle's LoS--NLoS transition. Matching the unshadowed two-state prior gain at $r_{\mathrm{ref}}=(55^2+h_{\mathrm{BS}}^2)^{1/2}$~m to $\beta_0(r_{\mathrm{ref}}/r_0)^{-\nu}$ sets $\beta_0^{\mathrm{LoS}}=\beta_0(r_{\mathrm{ref}}/r_0)^{-\nu}/\allowbreak[(1-\bar{p}_{\mathrm{blk}})+\allowbreak\bar{p}_{\mathrm{blk}}\kappa_{\mathrm{NLoS}}]$ with $\kappa_{\mathrm{NLoS}}=10^{-1.1}$ or $-11$~dB and $\bar{p}_{\mathrm{blk}}=\min\{0.9,\allowbreak(N_{\mathrm{obs}}/\pi)\allowbreak\arcsin((\bar{\varrho}_{\mathrm{obs}}+r_{\mathrm{map}})/\bar{d}_{\mathrm{obs}})\}=0.566$ at $(\bar{\varrho}_{\mathrm{obs}},\bar{d}_{\mathrm{obs}})=(4.3,28)$~m, or $44.6\%$ unpadded $55$~m-ring blockage. Using $(\nu_{\mathrm{LoS}},\nu_{\mathrm{NLoS}})=(2.4,3.5)$, $\beta_{\mathrm{LoS}}(r)=\beta_0^{\mathrm{LoS}}(r/r_{\mathrm{ref}})^{-\nu_{\mathrm{LoS}}}$, $\beta_{\mathrm{NLoS}}(r)=\kappa_{\mathrm{NLoS}}\beta_0^{\mathrm{LoS}}(r/r_{\mathrm{ref}})^{-\nu_{\mathrm{NLoS}}}$, and $48$-feature Fourier shadowing $s_{\mathrm{sh}}(x)=3\tanh(\tilde{s}(x)/3)$ for $\tilde{s}(x)=(2/48)^{1/2}\allowbreak\sum_{r=1}^{48}\cos((k_r^{\mathrm{sh}})^\top x+\allowbreak\varphi_r^{\mathrm{sh}})$ with $18$~m correlation and $\sigma_{\mathrm{sh}}=2.5$~dB, the smooth transmittance $T_{\mathrm{obs}}(x)$ and CKM gain $\bar{\beta}(x)$ in~\eqref{eq:channel} are
\begin{align*}
T_{\mathrm{obs}}(x)&=\prod_{o=1}^{N_{\mathrm{obs}}}\!\big[1-g_o(x)\,\sigma_{\mathrm{lg}}(-\Delta d_o(x)/\ell_{\mathrm{sm}})\big],\\
\bar{\beta}(x)&=\beta_{\mathrm{NLoS}}(r)\big[\beta_{\mathrm{LoS}}(r)/\beta_{\mathrm{NLoS}}(r)\big]^{T_{\mathrm{obs}}(x)}10^{\sigma_{\mathrm{sh}}s_{\mathrm{sh}}(x)/10}.
\end{align*}

On the $20$ rings at $D_i\in\{131,384\}$~kbits with $(\eta_{\mathrm{sec}},\eta_E)=(0.05,2\times 10^{-7})$, each scheme runs Algorithm~\ref{alg:proposed} at most once with $L_{\max}=14$ and infeasible starts scored zero over $512$ realizations under obstacle jitter $\Delta z_o^{\mathrm{obs}}\sim\mathcal{N}(0_2,\sigma_{\mathrm{map}}^2 I_2)$ at $\sigma_{\mathrm{map}}=0.20$~m and $32$-feature residual shadowing $\delta_{\mathrm{sh}}(x)$ with $\sigma_{\mathrm{err}}=1.0$~dB and $15$~m correlation. Segment--disk intersection on $[b,x_e^{\mathrm{tx}}]$ sets binary blockage in~\eqref{eq:rician_phy}, scaling $\beta_{\mathrm{LoS}}(r_e)$ at $K_{\mathrm{R}}=8$~dB and $\sigma_\theta=12^\circ$ on LoS rays or $\beta_{\mathrm{NLoS}}(r_e)$ at $K_{\mathrm{R}}=0$ and $\sigma_\theta=25^\circ$ on blocked rays by $10^{\sigma_{\mathrm{sh}}s_{\mathrm{sh}}(x_e^{\mathrm{tx}})/10}\allowbreak\cdot 10^{\sigma_{\mathrm{err}}\delta_{\mathrm{sh}}(x_e^{\mathrm{tx}})/10}$. Alongside ``CKM-Proposed,'' ``CKM-Serial,'' ``CKM-Co-located,'' ``CKM-Inner,'' and ``CKM-Fixed,'' feasible on $3/20$ rings at $384$~kbits, two baselines with $\bar{\beta}(x)=\beta_0(r/r_0)^{-\nu}$ run: ``Blind-Uncal'' keeps $\zeta=0.80$, whereas ``Blind-Margin'' sets $\zeta_{\mathrm{blind}}=\zeta\log_2(1+\allowbreak 10^{-M_{\mathrm{fade}}/10}\gamma_{\mathrm{ref}})/\allowbreak\log_2(1+\gamma_{\mathrm{ref}})=0.613$ using $\gamma_{\mathrm{ref}}=p_{\max}\beta_0(r_{\mathrm{ref}}/r_0)^{-\nu}/\allowbreak(N_0 W F_n)$ and fade margin $M_{\mathrm{fade}}=10\log_{10}(1/\kappa_{\mathrm{NLoS}})+\allowbreak 1.5\sigma_{\mathrm{sh}}=14.75$~dB.

In Table~\ref{tab:ckm_shadow}, ``Blind-Uncal'' suffers NLoS outage at $22.1\%$ and $13.8\%$ completion; at $131$~kbits, ``Blind-Margin'' and ``CKM-Proposed'' reach comparable utility, $0.2233$ versus $0.2224$ at Wilcoxon $p=0.50$, as light payloads absorb the $14.75$~dB margin. At $384$~kbits, ``CKM-Proposed'' improves utility by $24.6\%$ over ``Blind-Margin'' at $p=3.8\times 10^{-6}$, $8.7\%$ over ``CKM-Inner'' at $p=0.021$ and $3.0\%$ on $19$ shared-feasible rings, $12.5\%$ over ``CKM-Co-located'' at $p=3.1\times 10^{-4}$, and $4.6\%$ or $+0.0082$ over ``CKM-Serial'' with $95\%$ CI $[+0.0042,+0.0120]$ and $p=1.9\times 10^{-3}$. This is because ``Blind-Margin'' at $50.5\%$ completion and $47.9\%$ blockage and ``CKM-Inner'' with $19/20$ feasible rings and $47.0\%$ blockage stay exposed to radial shadows, whereas ``CKM-Proposed'' shifts $x_e^{\mathrm{tx}}$ by $1.68$~m on average and $1.22$~m transversely, cutting NLoS to $32.4\%$ while compressing during travel.

\section{Conclusion}
\label{sec:conclusion}

In this paper, we have investigated energy-efficient joint waypoint, semantic compression, power, and scheduling optimization for a multi-robot MEAN. Under exogenous latest-valid-packet estimation with calibrated symmetric conditionally independent errors, max--min quality--freshness utility maximization is equivalent to worst-target Bayesian risk minimization for a symmetric binary Markov source. For a fixed plan, we have developed a three-block BCD--SCA algorithm with a seven-regime compression KKT law at stationary anchors, monotonic convergence of retained iterates, analytical bounds, and a pipelined relocation threshold. Simulation results show that under equal battery budgets, the proposed algorithm improves energy efficiency with $28.2\%$ higher utility and $17.2\%$ lower energy than outer-boundary waypoints and $16.7\%$ below full-feature transmission, while outperforming co-located operation by $9.2\%$ under heavy payloads and fade-margin, co-located, and serial schemes by $24.6\%$, $12.5\%$, and $4.6\%$ under shadowing. Future work will address mobile targets and dynamic obstacles.

\appendices
\section{\texorpdfstring{Proof of Proposition~\ref{prop:kkt}}{Proof of Proposition 1}}
\label{app:proof_prop1}
Because $c_e$ enters~\eqref{eq:sub_Qt}--\eqref{eq:sub_Q} only with non-negative weights $\Psi_e^{\mathrm{cpu}}\ge 0$, eliminating $c_e=\Gamma_{i_e}(\rho_e)\ge 0$ directly in the convex primal subproblem decomposes the $(\boldsymbol{\rho},\mathbf{c})$ Lagrangian into separable convex costs $\sum_e J_e(\rho_e)$, giving~\eqref{eq:kkt_inc}. At a stationary anchor, $\eta_\rho(\rho_e^\star-\rho_e^b)=0$ and $M_e^{\mathrm{fid}}(\rho_e^\star)=\varpi_e\psi_{i_e}'(\rho_e^\star)$ reduce~\eqref{eq:kkt_inc} to $-\tilde{\vartheta}_e^0(\rho_e^\star)\in\partial\Gamma_{i_e}(\rho_e^\star)/c_{i_e}+N_{[\rho_{\min},1]}(\rho_e^\star)$ when $\Psi_e^{\mathrm{cpu}}>0$. Evaluating the piecewise-affine subdifferential at endpoints, open segments, and breakpoints $\{\varrho_1,\varrho_2\}$ with monotone $\tilde{\vartheta}_e^0$ yields~\eqref{eq:prop1_regimes}, where $\mathcal{S}_e^0$ is implicit for general $C^1$ concave $\psi_i$. For $\eta_\rho>0$, strong convexity on each $I_\varsigma$ gives unique $r_\varsigma$, closing to~\eqref{eq:prop1_proj} when $M_e^{\mathrm{fid}}\equiv 0$, with breakpoint condition $\rho_e^\star=\varrho_a\Leftrightarrow \varsigma_a c_{i_e}\Psi_e^{\mathrm{cpu}}\le\widetilde{\Phi}_e^{\mathrm{net}}(\varrho_a)+\eta_\rho(\varrho_a-\rho_e^b)\le \varsigma_{a+1}c_{i_e}\Psi_e^{\mathrm{cpu}}$. Finally, right-derivative optimality on strictly convex $J_e^0$ gives $\rho_e^\star>\varrho_a\Leftrightarrow(J_e^0)_+'(\varrho_a)<0\Leftrightarrow\psi_{i_e}'(\varrho_a)>(\vartheta_e-\varsigma_a)/\tilde{\lambda}_e$; at fixed duals, $\psi_i'(\varrho_a)$ for $\psi_i(\rho)=(1-2^{-\alpha_i\rho})/(1-2^{-\alpha_i})$ at $\varrho_a\ge 1/2$ strictly decreases from $1$ to $0$ in $\alpha_i>0$, defining a unique critical steepness whenever $0<(\vartheta_e-\varsigma_a)/\tilde{\lambda}_e<1$.

\section{\texorpdfstring{Proof of Proposition~\ref{prop:upper_bound}}{Proof of Proposition 3}}
\label{app:proof_prop3}
For the energy bound, $C_{-e}\succeq I_M$ gives $\gamma_{e\ell}^0\le p_e G_e(\mathbf{x}^{\mathrm{tx}})$ and $f_e-s_e\ge D_e/(\zeta R_e^{\mathrm{su}})$. For $p_{k_e}^{\mathrm{cir}},p_{\max,k_e}>0$, differentiating $(p_{k_e}^{\mathrm{cir}}+p/\eta_{k_e}^{\mathrm{PA}})/\ln(1+p G_e)$ yields $(u-1)e^{u-1}=e^{-1}(\eta_{k_e}^{\mathrm{PA}}p_{k_e}^{\mathrm{cir}}G_e-1)$ with $u=\ln(1+p G_e)$, giving unique optimal power $p_e^{\mathrm{ee}}(\mathbf{x}^{\mathrm{tx}})=\min\{p_{\max,k_e},\allowbreak[\exp(1+\allowbreak\mathcal{W}_0(e^{-1}(\eta_{k_e}^{\mathrm{PA}}p_{k_e}^{\mathrm{cir}}\allowbreak G_e(\mathbf{x}^{\mathrm{tx}})-1)))-1]/\allowbreak G_e(\mathbf{x}^{\mathrm{tx}})\}$ via principal Lambert-$W$ branch $\mathcal{W}_0$, approached as $p\downarrow 0$ if $p_{k_e}^{\mathrm{cir}}=0$. Piecewise-affine convexity places the packet energy minimum on $\mathcal{B}_{i_e}$, proving~\eqref{eq:bat_crit} and $D_{\mathrm{TD}}^{\mathrm{col}}(\bar{\rho})$ via $\sum_e(f_e-s_e)\le H$. For the spatial-DoF bound, let $V(t)=\sum_{e\in\mathcal{A}(t)}p_e h_e h_e^H$ with $\mathrm{rank}(V)\le M_{\mathrm{eff}}$; then $\sum_{e\in\mathcal{A}(t)}\log_2(1+\gamma_e^0)\le\log_2\det(I_M+V(t))\le M_{\mathrm{eff}}\log_2(1+\mathrm{tr}(V(t))/M_{\mathrm{eff}})$, and integrating over $[0,H]$ via Jensen's inequality yields~\eqref{eq:dof_cap}.

For the utility bounds, under homogeneous path loss $\|x_e^{\mathrm{tx}}-b\|\ge[R_{i_e}^{\mathrm{in}}-\tilde{\Delta}_e]_+$ with $\tilde{\Delta}_e=\min\{\|x_e^{\mathrm{tx}}-x_e^{\mathrm{s}}\|,R_{i_e}^{\mathrm{in}}\}$, whereas under a CKM $\bar{\beta}(x_e^{\mathrm{tx}})\le\bar{\beta}_{i_e}^{\mathrm{ub}}(\tilde{\Delta}_e)$ with $\tilde{\Delta}_e=\min\{\|x_e^{\mathrm{tx}}-x_e^{\mathrm{s}}\|,v_{\max}H\}$; both yield $d_e-g_e\ge\ell_{i_e}^{\mathrm{svc}}(\rho_e,\tilde{\Delta}_e)\ge\ell_{i_e}^\star>\ell_{0,i_e}$. At fixed $\mathbf{x}$, relaxing interference and battery limits implies $d_e-g_e\ge\underline{\ell}_e(\rho_e;\mathbf{x})$ and $d_{e_1}\ge\underline{d}_i(\mathbf{x})\ge\ell_i^\star$ for first valid $e_1\in\mathcal{D}_i$. Applying Jensen's inequality to concave $h\mapsto 1-e^{-h/\tau_i}$ over active durations summing to $H-d_{e_1}\le H-\underline{d}_i(\mathbf{x})\le H-\ell_i^\star$ with monotonicity in $n_i\le\bar{n}_i$ proves~\eqref{eq:prop2_bound}--\eqref{eq:system_ub}.
\IEEEtriggeratref{24}

\end{document}